\documentclass[journal]{IEEEtran}
\usepackage[normalem]{ulem}
\usepackage[noadjust]{cite}
\usepackage{bbm}
\usepackage{array}
\usepackage{dcolumn}
\usepackage{epsfig}
\usepackage[intlimits]{amsmath}
\usepackage{amsmath, amsfonts, yhmath, bm}
\usepackage{amssymb}
\usepackage{psfrag}
\usepackage{color,soul}
\usepackage{xcolor,cancel}
\usepackage[normalem]{ulem}
\usepackage{enumerate}
\usepackage{stackengine}
\usepackage[noadjust]{cite}
\usepackage{graphicx}
\usepackage{caption}
\usepackage{subcaption}
\usepackage{multirow}
\usepackage{cite}
\usepackage[font=footnotesize]{caption}
\usepackage{etoolbox}
\usepackage{tcolorbox}
\usepackage{float}
\usepackage[ruled,vlined]{algorithm2e}
\newcommand {\myvec}[1] {{\mbox{\boldmath $#1$}}}
\newcommand {\mymat}[1]  {{\mbox{\boldmath $#1$}}}
\newcommand{\bs}[1]{\boldsymbol{#1}}

\DeclareMathAlphabet      {\mathbfit}{OML}{cmm}{b}{it}

\newcommand{\etal}{\textit{et al.}}

\newcommand {\A} {\mymat{A}}

\newcommand {\mLambda} {\mymat{\Lambda}}

\newcommand {\C} {\mymat{C}}

\newcommand {\mPhi} {\mymat{\Phi}}
\newcommand {\tpsi} {\widetilde{\psi}}
\newcommand {\tupsi} {\widetilde{\upsi}}

\newcommand {\Ep} {\mymat{\mathcal{E}}}
\renewcommand {\H} {\mymat{H}}

\newcommand {\uvarep} {\myvec{\varepsilon}}
\newcommand {\dbaruvarep} {\overline{\overline{{\uvarep}}}}
\newcommand {\ueta} {\myvec{\eta}}
\renewcommand {\O} {\mathbf{O}}

\newcommand {\R} {\mymat{R}}

\newcommand {\hR} {\widehat{\R}}
\newcommand {\CR} {\bs{\mathcal{C}}}
\newcommand {\I} {\mymat{I}}
\newcommand {\X} {\mymat{X}}

\newcommand {\ue} {\myvec{e}}

\newcommand {\ua} {\myvec{a}}

\newcommand {\umu} {\myvec{\mu}}
\newcommand {\uhmu} {\widehat{\umu}}
\newcommand {\unu} {\myvec{\nu}}
\newcommand {\uhnu} {\widehat{\unu}}

\newcommand {\uxi} {\myvec{\xi}}

\newcommand {\hmLambda} {\widehat{\mLambda}}
\newcommand {\thmLambda} {\widehat{\widetilde{\mLambda}}}

\newcommand {\uepsilon} {\myvec{\epsilon}}
\newcommand {\dbaruepsilon} {\overline{\overline{{\uepsilon}}}}

\newcommand {\uc} {\myvec{c}}
\newcommand {\ux} {\myvec{x}}

\newcommand {\ur} {\myvec{r}}
\newcommand {\urho} {\myvec{\rho}}
\newcommand {\uiota} {\myvec{\iota}}

\newcommand {\ualpha} {\myvec{\alpha}}
\newcommand {\uv} {\myvec{v}}

\newcommand {\uo} {\myvec{0}}
\newcommand {\us} {\myvec{s}}

\newcommand {\uy} {\myvec{y}}
\newcommand {\uw} {\myvec{w}}

\newcommand {\uphi} {\myvec{\phi}}
\newcommand {\upsi} {\myvec{\psi}}

\newcommand {\utheta} {\myvec{\theta}}

\newcommand {\uon} {\myvec{1}}
\newcommand {\ep} {\mathcal{E}}

\newcommand {\Rset} {\mathbb{R}}
\newcommand {\Cset} {\mathbb{C}}

\newcommand {\Eset} {\mathbb{E}}
\newcommand {\Nset} {\mathbb{N}}
\newcommand {\mPsi} {\mymat{\Psi}}
\newcommand {\mSigma} {\mymat{\Sigma}}
\newcommand {\hmSigma} {\widehat{\mSigma}}
\newcommand {\Diag} {\text{\normalfont Diag}}

\newcommand {\tps} {\rm{T}}

\newcommand{\stkout}[1]{
	\color{red}\ifmmode\text{\sout{\ensuremath{#1}}}\else\sout{#1}\fi\color{black}}

\newcommand{\cmmnt}[1]{}

\newcommand{\addra}{}
\newcommand{\addraAQ}{}
\newcommand{\delra}{\cmmnt}
\newcommand{\delraAQ}{\cmmnt}
\newcommand{\delralign}{\cmmnt}

\begin{document}

\title{\addra{Asymptotically }Optimal Blind Calibration of Uniform Linear Sensor Arrays for Narrowband Gaussian Signals}

\author{Amir Weiss and Arie Yeredor
	
	\thanks{\delra{The authors are}\addra{A.\ Weiss is with the Department of Computer Science and Applied Mathematics, Faculty of Mathematics and Computer Science, Weizmann Institute of Science, 234 Herzl Street, Rehovot 7610001, Israel (e-mail: amir.weiss@weizmann.ac.il). A.\ Yeredor} with the School of Electrical Engineering, Faculty of Engineering, Tel-Aviv University,
		P.~O.~Box 39040, Tel-Aviv 69978, Israel\delra{,}\addra{ (e-mail:}
		\delra{amirwei2@mail.tau.ac.il, }arie@eng.tau.ac.il\addra{)}. Some parts of this work were published in our recent conference paper \cite{weiss2019optimalblind}.}
}

\maketitle

\begin{abstract}
An asymptotically optimal blind calibration scheme of uniform linear arrays for narrowband Gaussian signals is proposed. Rather than taking the direct Maximum Likelihood (ML) approach for joint estimation of all the unknown model parameters, which leads to a multi-dimensional optimization problem with no closed-form solution, we revisit Paulraj and Kailath's (P-K's) classical approach in exploiting the special (Toeplitz) structure of the observations' covariance. However, we offer a substantial improvement over P-K's ordinary Least Squares (LS) estimates by using asymptotic approximations in order to obtain simple, non-iterative, (quasi-)linear Optimally-Weighted LS (OWLS) estimates of the sensors\delra{'} gains and phases offsets with asymptotically optimal weighting, based only on the empirical covariance matrix of the measurements. Moreover, we prove that our resulting estimates are also asymptotically optimal w.r.t.\ the raw data, and can therefore be deemed equivalent to the ML Estimates (MLE), which are otherwise obtained by \textit{joint} ML estimation of \textit{all} the unknown model parameters. After deriving computationally convenient expressions of the respective Cram\'er-Rao lower bounds, we also show that our estimates offer improved performance when applied to non-Gaussian signals (and/or noise) as \delra{Q}\addra{q}uasi-MLE\delra{ (QMLE)} in a similar setting. The optimal performance of our estimates is demonstrated in simulation experiments, with a considerable improvement (reaching an order of magnitude and more) in the resulting mean squared errors w.r.t.\ P-K's ordinary LS estimates. We also demonstrate the improved accuracy in a multiple-sources directions-of-arrivals estimation task.
\end{abstract}


\section{Introduction}\label{sec:intro}

An obvious condition for the proper operation of sensor arrays in a variety of applications (e.g., beamforming or Direction of Arrival (DOA) estimation) is the precise calibration of their elements\delra{.}. Unfortunately, due to practical difficulties (e.g., \delra{sensor positioning errors}\addra{temperature variations or frequency drifts in the receivers}), errors in the model parameters, such as relative gain and phase variations within and among sensors, are present quite often. This, in turn, might translate into substantial degradation in the resulting performance. Therefore, the sensor array needs to be calibrated from time to time (if not upon each use).

While ``offline" calibration (i.e., prior to the ``operational" activity), using known calibration signals \addra{at known locations} when possible, is relatively simple, self or blind calibration \addra{(e.g., \cite{wijnholds2009multisource,balzano2007blind}) }is typically a more desirable, yet a more challenging task. In this paper, we address the blind calibration of the gains and phases in a sensor array within the framework of narrowband signals.\addra{ Naturally, the general blind calibration problem has already been widely addressed in the literature and is quite well-studied, as presented in the following survey.}

\addra{\subsection{Related Work}\label{subsec:relatedwork}
Blind calibration plays an important role in the overall successful operation in many applications. For example, in pushbroom cameras, where image destriping is necessary due to sensor-to-sensor variation within instruments, blind calibration was proposed in \cite{gadallah2000destriping} as an outliers-resilient alternative to histogram matching. In acoustic sensor arrays, blind calibration is attractive when both the sources and sensors locations are not known \textit{a-priori} \cite{moses2003self,raykar2004automatic} (see also \cite{weiss1989array} for this topic). Other examples may be found in the context of environmental sensor networks \cite{fishbain2016self,maag2018survey}, radio astronomy \cite{van2007self}, compressive-sensing-based imaging sensors \cite{cambareri2016non}, and timing offsets compensation of multi-channels analog-to-digital converter \cite{huang2007blind}, to name a few.

To address these problems, spanning over a wide variety of applications, several calibration models and assumptions were proposed. Classical calibration models consist of linear / affine \cite{balzano2007blind,lipor2014robust}, polynomial \cite{dorffer2016nonlinear} and multilinear \cite{fang2017using} relations between some (possibly unknown) input and output parameters, as well as low-rank and sparsity assumptions \cite{dorffer2016nonlinear,liu2013unified,bilen2014convex,liu2016sparse,chiarucci2018blind}.}

\delra{Naturally, this problem has already been widely addressed in the literature and is quite well-studied. }A few \addra{additional }important examples\addra{, which are more closely related to our work,} are Paulraj and Kailath's Least Squares (LS) (based) estimates for the unknown sensor gains and phases \cite{paulraj1985direction}, Friedlander and Weiss' eigenstructure method \cite{weiss1990eigenstructure}, which jointly calibrates the array and estimates the sources' DOAs, and the direct (rather involved) Maximum Likelihood (ML) approach, proposed in \cite{ng1996sensor} by Chong and See, accounting also for mutual coupling\addra{ (e.g., \cite{lin2006blind})} as well as for errors in the sensor positions, in which the ML Estimate (MLE) is pursued by an iterative algorithm (which does not necessarily converge to the MLE).\addra{ Viberg and Swindlehurst took a Bayesian approach in \cite{viberg1994bayesian}, where a maximum \textit{a-posteriori} estimate is proposed, assuming that certain prior knowledge of the array response errors is available.} More recent examples are due to Liu \etal's \cite{liu2010precision} and Wijnholds and Noorishad \cite{wijnholds2012statistically}, where a diagonal Weighted LS (WLS) and the weighted alternating LS estimates are proposed, resp. Nevertheless, these weighting approaches are essentially heuristic and are not shown (nor claimed) to be optimal\delra{ (in any sense)}.

\addra{\subsection{Merits and Contributions of this Work}\label{subsec:merits}}
In this paper we revisit the problem of blind sensor gains and phases estimation \addra{in Uniform Linear Arrays (ULAs) }for Gaussian signals, i.e., when the sources' DOAs and powers, as well as the noise level, are considered unknown. Following \cite{weiss2019optimalblind}, we extend our previous approach into a \textit{joint} estimation scheme of the gains and phases, and derive closed-form expressions of their \delra{(}approximate\delra{)} MLEs via Optimally-Weighted LS (OWLS) estimation. \addra{Despite previous claims regarding the (alleged) independence of the gain and phase estimation errors \cite{li2006theoretical}, we prove that these estimation errors are indeed correlated. Furthermore, w}\delra{W}e prove that the derived estimates asymptotically coincide with the MLEs in \textit{joint} estimation of \textit{all} the unknown model parameters w.r.t.\ the raw data\addra{ measurements}, and demonstrate this optimality in simulations.\addra{ Note that in this paper, by ``asymptotically", we refer to the case where \textit{only} the sample size $T$ approaches infinity. In particular, all our results are valid for any Signal-to-Noise Ratio (SNR) level, as long as it is fixed when increasing $T$. Furthermore, in this paper optimality is w.r.t. the minimal attainable MSE in unbiased estimation of the unknown deterministic parameters.}

The provided (non-iterative) solutions are efficiently computed, and as we demonstrate empirically in simulations, the resulting Mean Squared Errors (MSEs) are improved (in some scenarios) by more than an order of magnitude w.r.t.\ the MSEs attained by Paulraj and Kailath's ordinary LS estimates, and attain the performance bounds\addra{,} \delra{(}which are otherwise attained asymptotically by \textit{joint} ML estimation of \textit{all} the unknown model parameters\addra{ via non-convex, high-dimensional optimization}\delra{)}. In addition, we propose a \delra{(}simple\delra{)} generalization of the derived estimates for the non-Gaussian case, and show that while these generalized estimates are no longer optimal, they can still offer a significant improvement over the LS estimates.

The rest of the paper is organized as follows. The remainder of this section is devoted to a brief outline of our notations. In Section \ref{sec:problemformulation} we present the model under consideration and formulate the problem. Our asymptotically optimal blind calibration scheme is presented in Section \ref{sec:optimalcalibration}, where the OWLS estimates of the gains and phases are derived, and are shown (analytically) to be asymptotically the MLEs w.r.t.\ the raw data. In addition, simple \delra{(}approximated\delra{)} closed-form expressions of the Cram\'er-Rao Lower Bound (CRLB) on the MSEs are given as well, and the complementary Quasi-ML (QML) approach for non-Gaussian signals is briefly discussed in Subsection \ref{sec:QMLcalibration}. Simulations results, supporting our analytical derivations, are presented in Section \ref{sec:simulationresults}, and Section \ref{sec:conclusion} concludes the paper with final remarks.

{\subsection{Notations and Preliminaries}\label{subsec:notations}
We use $x, \ux$ and $\X$ for a scalar, column vector and matrix, resp. The superscripts $(\cdot)^{\tps}$, $(\cdot)^*$, $(\cdot)^{\dagger}$ and $(\cdot)^{-1}$ denote the transposition, complex conjugation, conjugate transposition and inverse operators, resp. We use $\I_{K}$ to denote the $K\times K$ identity matrix, and the pinning vector $\ue_k$ denotes the $k$-th column of the identity matrix with context-dependent dimension. Further, $\delta_{k\ell}\triangleq\ue_k^{\tps}\ue_{\ell}$ denotes the Kronecker delta of $k$ and $\ell$. We denote $\X_{k,:}\triangleq\ue_k^{\tps}\X$ (the $k$-th row of $\X$). $\Eset[\cdot]$ denotes expectation, the $\Diag(\cdot)$ operator forms an $M\times M$ diagonal matrix from its $M$-dimensional vector argument, and $\uo_M,\uon_M\in\Rset^{M\times 1}$ are the all-zeros and all-ones vectors, resp. We use $\jmath$ (a dotless $j$) to denote $\sqrt{-1}$; The operators $\Re\{\cdot\}$ and $\Im\{\cdot\}$ denote the real and imaginary parts (resp.) of their complex-valued argument.

\section{Problem Formulation}\label{sec:problemformulation}
Consider a ULA of $M$ sensors, each with an \textit{unknown} (deterministic) gain and phase response, and the presence of $N<M-1$ (unknown) narrowband sources \cite{astely1999spatial}, centered around some common carrier frequency with a wavelength $\lambda$, which are sufficiently far from the array to allow a planar wavefront (``far-field") approximation. Thus, let us denote the unknown gain and phase offset parameters as $\upsi\in\Rset_+^{M \times 1}$ and $\uphi\in[-\pi,\pi)^{M \times 1}$, resp., where $\psi_m$ and $\phi_m$ are the unknown gain and phase offsets of the $m$-th sensor, resp. 

More specifically, assuming that the received signals are Low-Pass Filtered (LPF)\footnote{The bandwidth of the LPF exceeds the bandwidth of the widest source.} and sampled at (at least) the Nyquist rate, \addra{following \cite{paulraj1985direction,li2006theoretical,ramamohan2018blind} with the same signal model used therein, }the vector of sampled (baseband) signals from all the $M$ sensors is given (for all $t\in\{1,\ldots,T\}$) by
\begin{equation}\label{modelequation}
\ur[t]=\mPsi\mPhi\left(\A(\ualpha)\us[t]+\uv[t]\right)\triangleq\mPsi\mPhi\ux[t]\in\Cset^{M\times1},
\end{equation}
where
\begin{enumerate}[(i)]
	\item $\mPsi\triangleq\Diag(\upsi)\in\Rset_+^{M\times M}$, $\mPhi\triangleq\Diag\left(e^{\jmath\uphi}\right)\in\Cset^{M\times M}$;
	\item $\us[t]\triangleq\left[s_1[t]\;\cdots\;s_N[t]\right]^{\tps}\in\Cset^{N\times1}$ denotes the vector of sources with wavenumber $k=2\pi/\lambda$, impinging on the array from (unknown) azimuth angles $\ualpha\triangleq\left[\alpha_1\;\cdots\;\alpha_N\right]^{\tps}\in\Rset^{N\times1}$;
	\item $\A(\ualpha)\triangleq\left[\ua(\alpha_1)\;\cdots\;\ua(\alpha_N) \right]\in\Cset^{M\times N}$ denotes the nominal array manifold matrix, with the steering vectors $\ua(\alpha_n)\triangleq\left[1\;e^{\jmath k \gamma\cos(\alpha_n)}\;\cdots\;e^{\jmath k(M-1)\gamma\cos(\alpha_n)}\right]^{\tps}\in\Cset^{M\times1}$ as its columns ($\gamma$ being the inter-element spacing);
	\item $\uv[t]\in\Cset^{M\times1}$ denotes an additive noise vector, \addra{modeling ambinet noise or ``interfering" signals, }assumed to be \delra{(both }spatially and temporally\delra{)} independent, identically distributed (i.i.d.) zero-mean circular Complex Normal (CN) \cite{loesch2013cramer} with a covariance matrix $\R_v\triangleq\Eset\left[\uv[t]\uv[t]^{\dagger}\right]=\sigma_v^2\I_M$, where $\sigma_v^2$ is considered unknown; and
	\item $\ux[t]$ denotes the signal that would have been received in the absence of gain or phase offsets, namely with $\mPsi=\mPhi=\I_M$.
\end{enumerate}
We also assume that the sources may be modeled as mutually uncorrelated random processes. Particularly, in this work, $\us[t]$ is considered as a (temporally) i.i.d.\ zero-mean circular CN vector process with an unknown diagonal covariance matrix $\R_s\triangleq\Eset\left[\us[t]\us[t]^{\dagger}\right]$. Furthermore, we assume $\us[t]$ and $\uv[t]$ are statistically independent. As a consequence, it follows that
\begin{equation}\label{CN_samples}
\ur[t]\sim \mathcal{CN}\left(\uo_M,\R\right), \forall t\in\{1,\ldots,T\},
\end{equation}
where\delralign{
\begin{gather}\label{covariance_of_r}
\R\triangleq\Eset\left[\ur[t]\ur[t]^{\dagger}\right]=\mPsi\mPhi\C\mPhi^*\mPsi\in\Cset^{M\times M},\\
\C\hspace{-0.05cm}\triangleq\hspace{-0.05cm}\Eset\left[\ux[t]\ux[t]^{\dagger}\right]\hspace{-0.05cm}=\hspace{-0.05cm}\A(\ualpha)\R_s\A^{\dagger}(\ualpha)+\sigma_v^2\I_M\in\Cset^{M\times M},\label{covariance_C}
\end{gather}}\addra{
\begin{align}\label{covariance_of_r}
\hspace{-0.075cm}\R&\triangleq\Eset\left[\ur[t]\ur[t]^{\dagger}\right]=\mPsi\mPhi\C\mPhi^*\mPsi\in\Cset^{M\times M},\\
\hspace{-0.075cm}\C\hspace{-0.05cm}&\triangleq\hspace{-0.05cm}\Eset\left[\ux[t]\ux[t]^{\dagger}\right]\hspace{-0.05cm}=\hspace{-0.05cm}\A(\ualpha)\R_s\A^{\dagger}(\ualpha)+\sigma_v^2\I_M\in\Cset^{M\times M},\label{covariance_C}
\end{align}}
and we have used $\mPsi^{\dagger}=\mPsi$ and $\mPhi^{\dagger}=\mPhi^*$.

The problem at hand can now be formulated as follows. \textit{Given the statistically independent measurements $\left\{\ur[t]\right\}_{t=1}^{T}$ whose (identical) distribution is prescribed by \eqref{CN_samples}, estimate the unknown (deterministic) parameters $\left\{\upsi,\uphi\right\}$.}

Notice that in this ``blind" setup, for this formulation, the other unknowns, namely $\ualpha, \sigma_v^2$ and the diagonal elements of $\R_s$ are considered as nuisance parameters. However, for other problems described by the same model, the parameters of interest, and accordingly the nuisance parameters, may be defined differently. For example, in the DOAs estimation problem\addra{ (e.g., \cite{cao2013doa})}, $\ualpha$ are the ``goal" estimands, whereas $\upsi, \uphi, \sigma_v^2$ and the diagonal elements of $\R_s$ are considered as nuisance parameters. Nevertheless, our goal here is to provide an \delra{(}asymptotically\delra{)} optimal estimation scheme for $\upsi$ and $\uphi$, based on the understanding that the measurements $\{\ux[t]\}$ of a perfectly calibrated sensor array would be preferable (in terms of the attainable performance) to $\{\ur[t]\}$ in other estimation problems described by this model.

\section{Approximate Optimal Blind Calibration}\label{sec:optimalcalibration}
We begin by recognizing that an \delra{(}asymptotically\delra{)} optimal solution to our problem would be obtained by joint ML estimation of $\upsi, \uphi, \ualpha, \sigma_n^2$ and the diagonal elements of $\R_s$, which \delra{(}asymptotically\delra{)} yields efficient estimates (\hspace{1sp}\cite{ra1922mathematical}) of $\upsi, \uphi$. However, since the derivation of the likelihood equations for this model is rather involved, which, at any rate, leads to a highly nonlinear system of equations, and since the sufficient statistic in this model is the sample covariance matrix of the measurements $\hR\triangleq\frac{1}{T}\sum_{t=1}^{T}{\ur[t]\ur[t]^{\dagger}}\in\Cset^{M\times M}$, we resort to \delra{(}approximated\delra{)} OWLS estimation of $\upsi, \uphi$ based (only) on $\hR$. This approach will lead to simple estimates, obtained as the solution of a linear system of equations, which will be shown to asymptotically coincide with the MLEs obtained via joint ML estimation of all the unknown parameters.

\subsection{Sensors' Gains and Phases Approximate OWLS Estimation}\label{subsec:OWLSestimation}
The proposed estimates we shall present are, in some sense, improved versions of the LS estimates proposed by Paulraj and Kailath \cite{paulraj1985direction} on the premises of the following observation. Since for a ULA the nominal array manifold matrix $\A(\ualpha)$ is a Vandermonde matrix (e.g., \cite{petersen2008matrix}) and all the signals involved are uncorrelated, the covariance matrix $\C$ (in \eqref{covariance_C}) of a perfectly calibrated array is a Toeplitz matrix (e.g., \cite{gray2006toeplitz}). Therefore, using the fact that (from \eqref{covariance_of_r}) $R_{ij}=C_{ij}\psi_i\psi_j\cdot e^{\jmath\left(\phi_i-\phi_j\right)}$, we have for
\begin{gather}\label{Log_equation}
\log\left(R_{ij}\right)\triangleq\mu_{ij}+\jmath\cdot\nu_{ij},
\end{gather}
the following relations
\begin{equation}\label{real_log_eq}
\mu_{ij}=\Re\left\{\log\left(R_{ij}\right)\right\}=\Re\left\{\log\left(c_{|i-j|+1}\right)\right\}+\log(\psi_i)+\log(\psi_j),
\end{equation}
\begin{equation}\label{imag_log_eq}
\nu_{ij}=\Im\left\{\log\left(R_{ij}\right)\right\}=\Im\left\{\log\left(c_{|i-j|+1}\right)\right\}+\phi_i-\phi_j,
\end{equation}
for any pair of indices $i,j\in\{1,\ldots,M\}$, where\delra{ $c_{|i-j|+1}\triangleq C_{ij}$ (since $C_{ij}=C_{k\ell}$ for any four indices satisfying $i-j=k-\ell$)}\addra{
\begin{equation}\label{definitionofthevectorc}
\begin{gathered}
c_{|i-j|+1}\triangleq C_{ij},\quad\forall i,j\in\{1,\ldots,M\}\\
\Longrightarrow\;\uc\triangleq\left[c_1\;\ldots\;c_{M}\right]^{\tps}\in\Cset^{M\times 1},
\end{gathered}
\end{equation}
since $C_{ij}=C_{k\ell}$ for any four indices satisfying $i-j=k-\ell$}. In particular, for any four indices satisfying $i-j=k-\ell$,\delra{ we also have}
\begin{equation}\label{log_LS_decoupled1}
\mu_{ij}-\mu_{k\ell}=\log(\psi_i)+\log(\psi_j)-\log(\psi_k)-\log(\psi_\ell),
\end{equation}
\begin{equation}\label{log_LS_decoupled2}
\nu_{ij}-\nu_{k\ell}=\phi_i-\phi_j-\phi_k+\phi_\ell.
\end{equation}
\addra{Note that for a unique definition of $\nu_{ij}$ in \eqref{Log_equation}, the linear relation in \eqref{imag_log_eq} can only hold when the result lies in the interval $[-\pi,\pi)$. Otherwise, a modulo operation is invoked, giving rise to a phase wrapping problem. However, in this work we assume that all the phase offsets are relatively ``small", i.e.,
\begin{equation}\label{phasescondition}
\left|\phi_{i}\right|\ll\pi,\quad\forall i\in\{1,\ldots,M\},
\end{equation}
such that \eqref{log_LS_decoupled2} surely holds, in contrast to \eqref{imag_log_eq}, which may be dominated by its first term. Note further that assumption \eqref{phasescondition} is considered standard, and is very reasonable in the context of array calibration errors, which is our main target in this work.}

Based on the relations \eqref{log_LS_decoupled1}--\eqref{log_LS_decoupled2}, and due to the fact that, in practice, the true covariance matrix $\R$ is not available, it was proposed in \cite{paulraj1985direction} to use $\hR$ instead of $\R$ and to collect all the nonredundant relations for which $(i,j)$ and $(k,\ell)$ pairs lie on the same main/super diagonals, and estimate, \textit{separately}, the gains and the phases using ordinary LS estimates which stem from the relations in \eqref{log_LS_decoupled1}--\eqref{log_LS_decoupled2} (see \cite{paulraj1985direction} for further details). 

Indeed, theoretically, $\hR$ can be made arbitrarily close to $\R$ by increasing (appropriately) the sample size $T$. However, in practice, the available sample size is always limited and is oftentimes fixed. Therefore, rather than relying on the coarse approximation $\hR\approx\R$, which leads, in this case, to the coarse, (sub-optimal) ordinary LS estimate, we propose a more refined analysis, which takes into account the estimation errors in $\hR$ and exploits (some of) their approximated statistical properties for obtaining a more accurate estimate, which will also be shown to be asymptotically the Uniformly Minimum-Variance Unbiased Estimate (UMVUE, \cite{lehmann2006theory}).

More formally, for any finite sample size $T$, we have
\begin{equation}\label{Restimate}
\hR\triangleq\R+\Ep\;\Rightarrow\;\widehat{R}_{ij}=R_{ij}+\ep_{ij}, \; \forall i,j\in\{1,\ldots,M\},
\end{equation}
where $\{\ep_{ij}\}$ denote the estimation errors in the estimation of $\{R_{ij}\}$. Hence, rewriting \eqref{Log_equation} with $\widehat{R}_{ij}$ replacing $R_{ij}$ yields
\begin{align}\label{log_equation_exact}
\hspace{-0.1cm}\log\left(\widehat{R}_{ij}\right)&\triangleq\widehat{\mu}_{ij}+\jmath\cdot\widehat{\nu}_{ij}=\log\left(R_{ij}\right)+\underbrace{\log\left(1+\frac{\ep_{ij}}{R_{ij}}\right)}_{\triangleq\zeta_{ij}}\\
&\triangleq(\mu_{ij}+\varepsilon_{ij})+\jmath\cdot(\nu_{ij}+\epsilon_{ij}),\label{log_equation_exact2}
\end{align}
for all $i,j\in\{1,\ldots,M\}$, where $\zeta_{ij}$ is the transformed (complex-valued) ``measurement noise", with $\varepsilon_{ij}$ and $\epsilon_{ij}$ as its real and imaginary parts, resp., such that we now have the following linear relations
\begin{equation}\label{real_exact_linear}
\widehat{\mu}_{ij}=\Re\left\{\log\left(c_{|i-j|+1}\right)\right\}+\log(\psi_i)+\log(\psi_j)+\varepsilon_{ij},
\end{equation}
\begin{equation}\label{imag_exact_linear}
\widehat{\nu}_{ij}=\Im\left\{\log\left(c_{|i-j|+1}\right)\right\}+\phi_i-\phi_j+\epsilon_{ij}.
\end{equation}
Combining the relations \eqref{log_LS_decoupled1}--\eqref{log_LS_decoupled2} and \eqref{real_exact_linear}--\eqref{imag_exact_linear}, one may obtain, again, two sets of linear equations, one for the unknowns $\{\log(\psi_m)\}$ only, and the other for $\{\phi_m\}$ only. Hence, the two resulting systems of linear equations, which now take into account the ``measurement" noise, are decoupled w.r.t.\ the unknowns $\{\psi_m\}$ and $\{\phi_m\}$. 

Ignoring other possible coupling, we have recently proposed \cite{weiss2019optimalblind} OWLS estimates of the gains and phases, which are based on the aforementioned two systems of linear equations and an independent, separate analysis of the transformed ``measurements" noise in each of these systems of equations. This means that each one of the proposed estimates in \cite{weiss2019optimalblind} is in fact optimal \textit{only} w.r.t.\ the statistics which appear in its corresponding system of equations, and not w.r.t.\ the full sufficient statistic $\hR$. For example, in \cite{weiss2019optimalblind}, as well as in Paulraj and Kailath \cite{paulraj1985direction}, the element $\widehat{R}_{1M}$ is discarded, although it is definitely a part of the full sufficient statistic $\hR$.

However, it turns out that although a deterministic decoupling is obtained using \eqref{log_LS_decoupled1}--\eqref{log_LS_decoupled2}, the two \addra{(exact) }systems of equations \addra{\eqref{real_exact_linear}--\eqref{imag_exact_linear} }are in fact \textit{statistically} coupled\addra{, as opposed to what is claimed in \cite{li2006theoretical}}. That is, the noise terms $\{\varepsilon_{ij}\}$ and $\{\epsilon_{ij}\}$ are correlated, meaning that more accurate estimates would be obtained by jointly estimating all the unknowns $\{\log(\psi_m),\phi_m,c_m\}$ via a widely linear estimate (e.g., \cite{picinbono1995widely}) based on all the complex measurements $\{\widehat{R}_{ij}\}$ and a unified, full analysis of the transformed ``measurement" noise $\{\zeta_{ij}\}$.

\delralign{To this end, let us denote $\tupsi\triangleq\log\left(\upsi\right)\in\Rset^{M\times 1}$ and $\log(\uc)\triangleq\urho+\jmath\cdot\uiota\in\Cset^{M\times 1}$, where $\log(\cdot)$ operates elementwise (notice that $c_1\in\Rset_+$), and define the vector of (real-valued) unknowns
\begin{equation}\label{vec_of_unknowns_old}
	\utheta \triangleq \left[\tupsi^{\tps} \;\, \uphi^{\tps} \;\, \urho^{\tps} \;\, \uiota^{\tps}\right]^{\tps}\in\Rset^{K_{\theta}\times 1},
\end{equation}
where $K_{\theta} \triangleq 4M-1\in\Nset$. With these notations, we may now write compactly
\begin{equation}\label{linear_LS_model_old}
	\uy=\H\utheta+\uxi\in\Rset^{M^2\times 1},
\end{equation}
where
\begin{align}\label{def_indices_equations1}
&\forall i\in\{1,\ldots,M\}:\forall j\in\{i,\ldots,M\}:\; y_m\triangleq\widehat{\mu}_{ij},\nonumber\\
&\H_{m,:}\triangleq\left[(\underbrace{\ue_i+\ue_j}_{\in\Rset^{M\times 1}})^{\tps} \; \uo^{\tps}_{M} \; \ue_{|i-j|+1}^{\tps} \; \uo^{\tps}_{M-1}\right]\in\Rset^{1\times K_{\theta}},\\
&(i,j)\leftrightarrow m\in\{1,\ldots,M(M+1)/2\},\nonumber
\end{align}
and
\begin{align}\label{def_indices_equations2}
&\forall i\in\{1,\ldots,M\}:\forall j\in\{i+1,\ldots,M\}:\; y_m\triangleq\widehat{\nu}_{ij},\nonumber\\
&\H_{m,:}\triangleq\left[\uo^{\tps}_{M} \; (\underbrace{\ue_i-\ue_j}_{\in\Rset^{M\times 1}})^{\tps} \; \uo^{\tps}_{M} \; \ue_{|i-j|+1}^{\tps}\right]\in\Rset^{1\times K_{\theta}},\\
&(i,j)\leftrightarrow m\in\{M(M+1)/2+1,\ldots,M^2\},\nonumber
\end{align}
such that the mapping $m\leftrightarrow(i,j)$ is injective and we have defined $\uxi\triangleq\left[\uvarep^{\tps} \;\, \uepsilon^{\tps}\right]^{\tps}\in\Rset^{M^2\times 1}$ which collects all the corresponding elements $\{\varepsilon_{ij}\}$ and $\{\epsilon_{ij}\}$ in compliance with \eqref{def_indices_equations1}--\eqref{def_indices_equations2}. Notice that the inherent ``blindness" of this formulation inflicts rank-deficiency on $\H$, which in turn implies non-identifiability of the gains and phases. Indeed, the observations \eqref{linear_LS_model} may be equivalently ``explained" by more than one estimate. Fortunately, this is also intuitively expected, since we can obviously only measure phase differences between different elements, the gain is well-defined only with at least one fixed power-related parameter, and in this blind scenario (where both $\uphi$ and $\ualpha$ are unknown) the spatial frequencies\footnote{The $n$-th spatial frequency is defined as $\omega_n \triangleq k \gamma\cos(\alpha_n)$.}, corresponding to the DOAs, may be determined only up to an arbitrary rotation (e.g., \cite{astely1999spatial}). Therefore, w.l.o.g., we add the following three ``references" equations to \eqref{linear_LS_model}
\begin{align}\label{ref_equations1}
&\tpsi_1=0\;\Leftrightarrow\;y_{M^2+1}=0,\;\H_{M^2+1,:}\triangleq\left[1 \; \uo^{\tps}_{K_{\theta}-1}\right],\\
&\phi_1=0\;\Leftrightarrow\;y_{M^2+2}=0,\;\H_{M^2+2,:}\triangleq\left[\uo^{\tps}_{M} \; 1 \; \uo^{\tps}_{K_{\theta}-M-1}\right],\label{ref_equations2}\\
&\phi_2=0\;\Leftrightarrow\;y_{M^2+3}=0,\;\H_{M^2+3,:}\triangleq\left[\uo^{\tps}_{M+1} \; 1 \; \uo^{\tps}_{K_{\theta}-M-2}\right],\label{ref_equations3}
\end{align}
(where $\tpsi_1=0\Leftrightarrow\psi_1=1$) with which $\H$ is now full-rank and accordingly the model is identifiable.}

\addra{To this end, let us denote $\tupsi\triangleq\log\left(\upsi\right)\in\Rset^{M\times 1}$ and $\log(\uc)\triangleq\urho+\jmath\cdot\uiota\in\Cset^{M\times 1}$, where $\log(\cdot)$ operates elementwise, and define the vector of (real-valued) unknowns
\begin{equation}\label{vec_of_unknowns}
\utheta \triangleq \left[\tupsi^{\tps} \;\, \uphi^{\tps} \;\, \urho^{\tps} \;\, \uiota^{\tps}\right]^{\tps}\in\Rset^{K_{\theta}\times 1},
\end{equation}
where $K_{\theta} \triangleq 4M$. With these notations, noting that according to \eqref{real_exact_linear} and \eqref{imag_exact_linear}, each element of the $M\times M$ Hermitian matrix $\log\left(\hR\right)$ is a linear combination of elements of $\utheta$ and additional noise terms, we may compactly write a linear ``correlation measurements" model
\begin{equation}\label{linear_LS_model}
\uy=\H\utheta+\uxi\in\Rset^{M^2\times 1},
\end{equation}
consisting of the ``non-replicated" real and imaginary parts of $\log\left(\hR\right)$. That is, the $M^2$ ``measurements" in $\uy$ consist of $0.5M(M+1)$ values of $\widehat{\mu}_{ij}$ for $(i,j)\in\{1,\ldots,M\}$ with $j\leq i$, and of $0.5M(M-1)$ values of $\widehat{\nu}_{ij}$ for $(i,j)\in\{1,\ldots,M\}$ with $j>i$. Likewise, the ``noise" vector $\uxi\triangleq\left[\uvarep^{\tps} \;\, \uepsilon^{\tps}\right]^{\tps}\in\Rset^{M^2\times 1}$ consists of the respective $0.5M(M+1)$ elements of the $\widehat{\mu}_{ij}$-related noise $\varepsilon_{ij}$ \eqref{real_exact_linear} (in $\uvarep$) and of the $0.5M(M-1)$ elements of the $\widehat{\nu}_{ij}$-related noise $\epsilon_{ij}$ \eqref{imag_exact_linear} (in $\uepsilon$). See Appendix \ref{AppA0} for the explicit structure of $\uy, \H$ and $\uxi$.

Notice that the inherent ``blindness" of this formulation inflicts rank-deficiency on $\H$, which in turn implies non-identifiability of the gains and phases. Indeed, the ``correlation measurements" $\uy$ may be equivalently ``explained" by more than one estimate. Fortunately, this is also intuitively expected, since: (i) we can obviously only measure phase differences between different elements; (ii) the gain is well-defined only with at least one fixed power-related parameter; and (iii) in this blind scenario (where both $\uphi$ and $\ualpha$ are unknown) the spatial frequencies\footnote{The $n$-th spatial frequency is defined as $\omega_n \triangleq k \gamma\cos(\alpha_n)$.} corresponding to the DOAs may be determined only up to an arbitrary rotation (e.g., \cite{astely1999spatial}). Therefore, w.l.o.g.\ we may arbitrarily set $\tpsi_1, \phi_1$ and $\phi_2$ to zero. Note also, that since $c_1$ is known to be real-valued (and positive), we also have $\iota_1=0$. We may therefore eliminate these parameters from $\utheta$, together with the four corresponding columns (the $1^{\text{st}}, M+1, M+2$ and $3M+1$) of $\H$, maintaining the same relation \eqref{linear_LS_model} with the newly defined $\utheta\in\Rset^{K_{\theta}\times 1}$ and $\H\in\Rset^{M^2\times K_{\theta}}$, only now $K_{\theta}=4M-4$, so that now $\H$ is full-rank and the (reduced) model is identifiable.}

Now, from the Gauss-Markov theorem \cite{lehmann2006theory}, the Best Linear Unbiased Estimate (BLUE) of $\utheta$ \delra{based on}\addra{given} $\uy$ is \delra{given by }the OWLS estimate
\begin{equation}\label{WLSestimate}
\widehat{\utheta}_{\tiny{\text{OWLS}}}\triangleq\left(\H^{\tps}\mLambda_{\xi}^{-1}\H\right)^{-1}\H^{\tps}\mLambda_{\xi}^{-1}\left(\uy-\ueta_{\xi}\right),
\end{equation}
where $\ueta_{\xi}\triangleq\Eset\left[\uxi\right]$ and $\mLambda_{\xi}\triangleq\Eset\left[\left(\uxi-\ueta_{\xi}\right)\left(\uxi-\ueta_{\xi}\right)^{\tps}\right]$ are the mean and covariance matrix of $\uxi$, resp. The BLUE attains the minimal attainable MSE matrix among all linear unbiased estimates, and when $\uxi$ is Gaussian, it is also the MLE of $\utheta$ (based on $\uy$), which is an efficient estimate (\hspace{1sp}\cite{ra1922mathematical}, even non-asymptotically), and therefore is also the UMVUE of $\utheta$ based on $\uy$.

Thus, our goal now is to obtain closed-form expressions (possibly approximated) for $\ueta_{\xi}$\delra{ and }\addra{, }$\mLambda_{\xi}$ in terms of the available and/or estimable quantities, in order to eventually obtain the estimate $\eqref{WLSestimate}$, or at least a well-approximated version thereof.\delra{ We note that in the following analysis, we shall refer only to the non-degenerate $M^2$ noise elements in \eqref{linear_LS_model}, although ultimately, due to \eqref{ref_equations1}--\eqref{ref_equations3}, $\uy$ is $(M^2+3)$-dimensional}\delralign{\footnote{\delra{In practice, for the computation of \eqref{MLbasedWLSestimate}, we treat the extra $3$ reference equations as uncorrelated, ``almost" noiseless ones. That is, their covariances with all $\{\varepsilon_{ij},\epsilon_{ij}\}$ are zero and their variances are set (arbitrarily) to be ``very small" values in comparison with $\tfrac{1}{T}$ (e.g., for $T=10^3$, we set $10^{-15}$)}.}.}

To this end, assume that $T$ is sufficiently large such that $|\ep_{ij}|\ll |R_{ij}|$ for all possible $(i,j)$. With this, using the second-order Taylor expansion approximation \delra{$\log(1+z)\approx z-\tfrac{z^2}{2}$ (which holds for all $z\in\Cset$ satisfying $|z|\ll 1$), }\addra{
\begin{equation}\label{logapproximation}
|z|\ll 1: \log(1+z)\approx z-\tfrac{z^2}{2}, \quad\forall z\in\Cset,
\end{equation}
}the equivalent ``measurement noise" $\zeta_{ij}$ reads
\begin{equation}\label{approximated_noise}
\zeta_{ij}=\log\left(1+\frac{\ep_{ij}}{R_{ij}}\right)\approx\frac{\ep_{ij}}{R_{ij}} - \frac{\ep_{ij}^2}{2R_{ij}^2}, \; \forall i,j\in\{1,\ldots,M\}.
\end{equation}
\addra{Recalling that $\hR$ is unbiased and u}\delra{U}sing the pseudo-covariance of $\ep_{ij}$ (derived in Appendix \ref{AppA}, see \eqref{errorcovariance}), \delra{and recalling that $\hR$ is unbiased, which implies that $\ep_{ij}$ has zero-mean, }\addra{we get
\begin{equation}\label{FirstandSecondOrderStatistics}
\Eset\left[\ep_{ij}\right]=0,\quad\Eset\left[\ep_{ij}^2\right]=\frac{1}{T}R^2_{ij},\quad\forall i,j\in\{1,\ldots,M\}.
\end{equation}
Therefore, }it follows that
\begin{align}\label{noiseexpectation}
\Eset\left[\zeta_{ij}\right]&\approx\frac{\Eset\left[\ep_{ij}\right]}{R_{ij}} - \frac{\Eset\left[\ep_{ij}^2\right]}{2R_{ij}^2}=-\frac{1}{2T}\;\delra{\Rightarrow}\addra{\quad\Longrightarrow}\\
\Eset\left[\varepsilon_{ij}\right]&\approx-\frac{1}{2T},\;\Eset\left[\epsilon_{ij}\right]\approx0,\;\forall i,j\in\{1,\ldots,M\},
\end{align}
so that 
\begin{equation}\label{noise_approx_mean}
\ueta_{\xi}\approx-\frac{1}{2T}\cdot\left[\uon^{\tps}_{0.5M(M+1)}\; \uo^{\tps}_{0.5M(M-1)}\right]^{\tps}\triangleq\widehat{\ueta}_{\xi}.
\end{equation}
As for the covariance matrix of $\uxi$, which also reads $\mLambda_{\xi}=\Eset\left[\uxi\uxi^{\tps}\right]-\ueta_{\xi}\ueta_{\xi}^{\tps}$, based on the assumption \eqref{CN_samples} that $\{\ur[t]\}$ are all circular CN, and in particular using Isserlis' theorem \cite{isserlis1918formula}, we show in Appendix \ref{AppA} that the elements of $\Eset\left[\uxi\uxi^{\tps}\right]$ are \delra{(}approximately\delra{)} given by
\begin{align}\label{approximated_covariance_noise}
&\forall i,k\in\{1,\ldots,M\}:\forall j,\ell\in\{i,\ldots,M\}:\nonumber\\
&\Eset\left[\varepsilon_{ij}\cdot\varepsilon_{k\ell}\right]\approx\frac{1}{T}\cdot0.5\cdot\Re\left\{\frac{R_{ik}R^*_{j\ell}}{R_{ij}R^*_{k\ell}}+\frac{R_{i\ell}R^*_{jk}}{R_{ij}R_{k\ell}}\right\},\\
&\Eset\left[\epsilon_{ij}\cdot\epsilon_{k\ell}\right]\approx\frac{1}{T}\cdot0.5\cdot\Re\left\{\frac{R_{ik}R^*_{j\ell}}{R_{ij}R^*_{k\ell}}-\frac{R_{i\ell}R^*_{jk}}{R_{ij}R_{k\ell}}\right\},\\
&\Eset\left[\varepsilon_{ij}\cdot\epsilon_{k\ell}\right]\approx\frac{1}{T}\cdot0.5\cdot\Im\left\{\frac{R_{i\ell}R^*_{jk}}{R_{ij}R_{k\ell}}-\frac{R_{ik}R^*_{j\ell}}{R_{ij}R^*_{k\ell}}\right\},\label{approximated_covariance_noise_last}
\end{align}
so that $\mLambda_{\xi}$ is \delra{(}approximately\delra{)} a function of $\R$ \textit{only}.\addra{ Note that these approximations, which are based on the approximation \eqref{logapproximation}, as well as on the asymptotic (complex) Normality of the estimation errors $\{\ep_{ij}\}$, become arbitrarily close for a sufficiently large $T$. Particularly, this holds for any fixed SNR. Note further that \eqref{approximated_covariance_noise_last} shows that $\varepsilon_{ij},\epsilon_{k\ell}$ are indeed dependent.}

Of course, the true $\R$ is in fact unknown. However, since $\hR$ is the MLE of $\R$, by virtue of the invariance property of the MLE \cite{mukhopadhyay2000probability}, it follows that $\hmLambda_{\xi}$, a matrix whose elements are computed by \eqref{noise_approx_mean} and \eqref{approximated_covariance_noise}--\eqref{approximated_covariance_noise_last}, but with $\{\widehat{R}_{ij}\}$ replacing $\{R_{ij}\}$, is \delra{(}approximately\delra{)} the MLE of $\mLambda_{\xi}$. Therefore, we propose the following ``ML-based OWLS" estimate
\begin{gather}\label{MLbasedWLSestimate}
\widehat{\utheta}_{\tiny{\text{ML-OWLS}}}\triangleq\left(\H^{\tps}\hmLambda_{\xi}^{-1}\H\right)^{-1}\H^{\tps}\hmLambda_{\xi}^{-1}\left(\uy-\widehat{\ueta}_{\xi}\right),
\end{gather}
from which the desired ML-based OWLS estimates of the gains and phases (for all $m\in\{1,\ldots,M\}$)
\begin{align}
&\left(\widehat{\psi}_m\right)_{\tiny{\text{ML-OWLS}}}=\exp\left(\ue_m^{\tps}\widehat{\utheta}_{\tiny{\text{ML-OWLS}}}\right),\label{OWLS_estimates1}\\
&\left(\widehat{\phi}_m\right)_{\tiny{\text{ML-OWLS}}}=\ue_{(M+m)}^{\tps}\widehat{\utheta}_{\tiny{\text{ML-OWLS}}},\label{OWLS_estimates2}
\end{align}
are readily extracted. \addra{Note that the inverse matrix $\left(\H^{\tps}\hmLambda_{\xi}^{-1}\H\right)^{-1}$ exists only when the inverse matrix $\hmLambda_{\xi}^{-1}$ exists, which is guaranteed (almost surely) when $T>M^2$, dictating the minimal sample size required for the validity of \eqref{MLbasedWLSestimate}, and hence \eqref{OWLS_estimates1}, \eqref{OWLS_estimates2}. Further, n}\delra{N}ote that for a sufficiently large $T$: $\widehat{\utheta}_{\tiny{\text{ML-OWLS}}}\approx\widehat{\utheta}_{\tiny{\text{OWLS}}}$
by virtue of the continuous mapping theorem \cite{mann1943stochastic} and the consistency of the MLE (\hspace{1sp}\cite{cramer2016mathematical}) $\hmLambda_{\xi}$.

\subsection{Approximate ML Estimation and Cram\'er-Rao Bound}\label{subsec:MLEestimation}
Since the ML estimation errors $\{\ep_{ij}\}$ are asymptotically (non-circular) jointly CN, the transformed estimation errors $\{\zeta_{ij}\}$, which can be asymptotically linearized by neglecting the quadratic term in \eqref{approximated_noise}, become \delra{(}asymptotically\delra{)} approximately jointly CN as well. Thus, it follows that $\widehat{\utheta}_{\tiny{\text{OWLS}}}$ is \delra{(}approximately\delra{)} also the MLE of $\utheta$ based on $\uy$, which means, in particular, that $\widehat{\uphi}_{\tiny{\text{OWLS}}}$ is \delra{(}approximately\delra{)} the MLE of $\uphi$ based on $\uy$. As for the gains estimates, which are obtained by (elementwise) exponentiation of $\widehat{\tupsi}_{\tiny{\text{OWLS}}}$, once again, using the invariance property of the MLE, it follows that $\widehat{\upsi}_{\tiny{\text{OWLS}}}$ is the MLE of $\upsi$ based on $\uy$. Therefore, we conclude that $\widehat{\uphi}_{\tiny{\text{ML-OWLS}}}$, $\widehat{\upsi}_{\tiny{\text{ML-OWLS}}}$ asymptotically coincide with the MLEs of $\uphi$, $\upsi$ (resp.) based on $\uy$. However, observe that $\uy$ is an invertible function of $\hR$. Consequently, since $\hR$ is the (minimal) sufficient statistic of model \eqref{modelequation}, if follows that $\widehat{\uphi}_{\tiny{\text{OWLS}}}$, $\widehat{\upsi}_{\tiny{\text{OWLS}}}$ are also asymptotically the MLEs of $\uphi$, $\upsi$ (resp.) based on the raw data $\{\ur[t]\}_{t=1}^T$. Hence, we conclude that $\widehat{\uphi}_{\tiny{\text{ML-OWLS}}}$, $\widehat{\upsi}_{\tiny{\text{ML-OWLS}}}$ are asymptotically the MLEs of $\uphi$, $\upsi$ (resp.) based on the raw data $\{\ur[t]\}_{t=1}^T$, and accordingly are (only, \cite{8423486}) asymptotically efficient. Note that while, asymptotically, $\widehat{\uphi}_{\tiny{\text{ML-OWLS}}}$ is also the UMVUE of $\uphi$, the fact that $\widehat{\tupsi}_{\tiny{\text{ML-OWLS}}}$ is the UMVUE of $\tupsi$ does not imply, in general, that $\widehat{\upsi}_{\tiny{\text{ML-OWLS}}}$ (still being the MLE of $\upsi$) is also the UMVUE of $\upsi$.

Clearly, using $\widehat{\mLambda}_{\xi}\approx\mLambda_{\xi}$, we have that
\begin{equation}\label{approx_MLE_covariance}
\Eset\left[\left(\widehat{\utheta}_{\tiny{\text{ML-OWLS}}}-\utheta\right)\left(\widehat{\utheta}_{\tiny{\text{ML-OWLS}}}-\utheta\right)^{\tps}\right]\approx\left(\H^{\tps}\mLambda_{\xi}^{-1}\H\right)^{-1}.
\end{equation}
But since we have shown that $\widehat{\utheta}_{\tiny{\text{ML-OWLS}}}$ is asymptotically the MLE of $\utheta$ w.r.t.\ the raw data $\{\ur[t]\}_{t=1}^T$, using the notation
\begin{equation}\label{MSE_matrix_notations}
\left(\H^{\tps}\mLambda_{\xi}^{-1}\H\right)^{-1}\triangleq \begin{bmatrix}
\CR_{\tpsi} & \CR_{\tpsi\phi} & \CR_{\tpsi\rho} & \CR_{\tpsi\iota}\\
\CR_{\tpsi\phi}^{\tps} & \CR_{\phi} & \CR_{\phi\rho} & \CR_{\phi\iota}\\
\CR_{\tpsi\rho}^{\tps} & \CR_{\phi\rho}^{\tps} & \CR_{\rho} & \CR_{\rho\iota}\\
\CR_{\tpsi\iota}^{\tps} & \CR_{\phi\iota}^{\tps} & \CR_{\rho\iota}^{\tps} & \CR_{\iota} \end{bmatrix} \in \Rset^{K_{\theta} \times K_{\theta}},
\end{equation}
we may conclude that the CRLBs on the MSEs of any unbiased estimate of the sensors\delra{'} gains and phases are given approximately by
\delralign{\begin{align}
\text{CRLB}(\psi_n)&\approx\left(\frac{\partial\upsi}{\partial\tupsi}\left(\CR_{\tpsi}\frac{\partial\upsi}{\partial\tupsi}\right)^{\tps}\right)_{nn}=e^{2\tpsi_n}\left(\CR_{\tpsi}\right)_{nn}\nonumber\\
&=\psi_n^2\left(\CR_{\tpsi}\right)_{nn},\quad\quad\forall n\in\{2,\ldots,M\},\label{approx_MLE_gains_covariance_old}\\
\text{CRLB}(\phi_m)&\approx\left(\CR_{\phi}\right)_{mm},\quad\quad\quad\forall m\in\{3,\ldots,M\}\label{approx_MLE_phases_covariance_old},
\end{align}}
\begin{align}
\text{CRLB}(\psi_n)&\approx\left(\frac{\partial\upsi}{\partial\tupsi}\,\CR_{\tpsi}\,\frac{\partial\upsi}{\partial\tupsi}^{\tps}\right)_{nn}=e^{2\tpsi_n}\left(\CR_{\tpsi}\right)_{nn}\nonumber\\
&=\psi_n^2\left(\CR_{\tpsi}\right)_{nn},\quad\quad\forall n\in\{2,\ldots,M\},\label{approx_MLE_gains_covariance}\\
\text{CRLB}(\phi_m)&\approx\left(\CR_{\phi}\right)_{mm},\quad\quad\quad\forall m\in\{3,\ldots,M\}\label{approx_MLE_phases_covariance},
\end{align}
where we have used $\partial\psi_m/\partial\tpsi_n=\delta_{mn}\cdot e^{\tpsi_n}=\delta_{mn}\cdot\psi_n$.

We note that the expressions given in \eqref{approx_MLE_gains_covariance}--\eqref{approx_MLE_phases_covariance} for the CRLB of $\upsi,\uphi$ are \addra{valid under the same conditions specified ear\addraAQ{li}er, after \eqref{approximated_covariance_noise_last}. In addition, these expressions are }somewhat less involved and require simpler computations than the ones which are obtained by direct computation of all the Fisher Information Matrix' (FIM) elements for \textit{all} the unknown parameters, followed by an inversion of the FIM.\addra{ Further, it is easy to verify that $\CR_{\tpsi\phi}$ is generally not all zeros, hence proving that the gain and phase estimation errors are \textit{not} independent.}

\subsection{Approximate QML-based Blind Calibration}\label{sec:QMLcalibration}
Consider the same received-signals model \eqref{modelequation} (Section \ref{sec:problemformulation}), with the Gaussianity assumption relaxed, i.e., the source signals $\us[t]$ and the additive noise $\uv[t]$ are only assumed (each) to be temporally i.i.d.\ proper (\hspace{1sp}\cite{neeser1993proper}) complex-valued random processes (with arbitrary probability distributions) and mutually uncorrelated.

In this general framework as well, as long as the fourth-order joint cumulants of the measurements are finite. i.e.,
\begin{multline}
\exists \varrho\in\Rset_+: \forall i,j,k,\ell\in\{1,\ldots,M\}:\\
|\kappa_r[i,j,k,\ell]|\triangleq|\text{cum}(r_i[t],r_j^*[t],r_k[t],r^*_\ell[t])|<\varrho,
\end{multline}
where $\text{cum}(r_i[t],r_j^*[t],r_k[t],r^*_\ell[t])$ denotes the fourth-order joint cumulant of its arguments, the estimate $\hR$, which is no longer necessarily the MLE of $\R$, is still consistent by virtue of the law of large numbers \cite{ross2009first}. Therefore, the proposed estimate \eqref{MLbasedWLSestimate} retains its consistency property, even for non-Gaussian signals $\ur[t]$, where the non-Gaussianity may be due to the sources' and/or the noise's distributions. However, in order to retain its asymptotic optimality w.r.t.\ the statistic $\hR$ (but now certainly not necessarily w.r.t.\ the raw data $\{\ur[t]\}_{t=1}^T$) even for non-Gaussian signals, the weight matrix $\widehat{\mLambda}_{\xi}^{-1}$ needs to be updated accordingly.

Thus, from the analysis presented in Appendix \ref{AppA}, in the general (not necessarily Gaussian) case, it follows that
\begin{align}
\Eset\left[\ep_{ij}\ep_{k\ell}^*\right]&=\frac{1}{T}\left(\kappa_r[i,j,\ell,k]+R_{ik}R^*_{j\ell}\right),\label{errorcovariance_general1}\\
\Eset\left[\ep_{ij}\ep_{k\ell}\right]&=\frac{1}{T}\left(\kappa_r[i,j,k,\ell]+R_{i\ell}R^*_{jk}\right).\label{errorcovariance_general2}
\end{align}
Continuing the same derivation in Appendix \ref{AppA} with the general expressions \eqref{errorcovariance_general1}--\eqref{errorcovariance_general2} yields the updated elements of $\mLambda_{\xi}$ for the general case. Obviously, when $\kappa_r[i,j,k,\ell]=0$ for all $i,j,k,\ell$, these expressions reduce back to \eqref{approximated_covariance_noise}--\eqref{approximated_covariance_noise_last}.

Of course, in the general case $\mLambda_{\xi}$ is not (even approximately) a function of $\R$ only, as it depends on terms which are determined by the fourth-order statistics of the received signals. Nevertheless, one may still construct a consistent estimate $\widehat{\mLambda}_{\xi}$, by replacing all $\{R_{ij}\}$ with $\{\widehat{R}_{ij}\}$ and all $\{\kappa_r[i,j,k,\ell]\}$ with some consistent estimates $\{\widehat{\kappa}_r[i,j,k,\ell]\}$ thereof (based on the available data $\{\ur[t]\}_{t=1}^T$). With this, the right-hand side of \eqref{MLbasedWLSestimate} becomes the ``QML-based OWLS", denoted by $\widehat{\utheta}_{\tiny{\text{QML-OWLS}}}$, since it can be viewed as the OWLS estimate which is based on $\hR$, the \textit{quasi} MLE of $\R$. This estimate still \delra{(}approximately\delra{)} attains the MSE matrix given in \eqref{approx_MLE_covariance}, which, in this case, no longer serves as the CRLB on the corresponding MSEs w.r.t.\ the raw data.

We note in passing that another plausible approach is to still construct the weight matrix $\widehat{\mLambda}_{\xi}^{-1}$ according to \eqref{noise_approx_mean} and \eqref{approximated_covariance_noise}--\eqref{approximated_covariance_noise_last}, and simply ignore the contributions of the fourth-order cumulants. In this case, $\widehat{\mLambda}_{\xi}$ would serve as an inconsistent (biased) estimate of $\mLambda_{\xi}$, and consequently the resulting estimate will no longer be (even asymptotically) the OWLS, but only some ``reasonable" WLS estimate. However, this estimate is considerably cheaper in terms of computational complexity, since it requires the terms of $\hR$ only and does not require estimation of fourth-order cumulants. Obviously, this comes at the cost of a compromised MSE, which nonetheless has a smaller constant gap from optimality (in [dB]) than the ordinary LS estimate. The (rather technical) analysis of this approximate QML-based estimate is out of the scope of this paper, and is therefore omitted.\addra{ Finally, summarizing this section, Algorithm \ref{Algorithm1} briefly describes the steps of our proposed blind (Q)ML-OWLS calibration scheme.
\setlength{\textfloatsep}{10pt}
\begin{algorithm}[t]
\addra{\KwIn{$\{\ur[t]\}_{t=1}^T$ (measured signals)}
		\KwOut{$\{\widehat{\ur}[t]\}_{t=1}^T$ (post-calibration measured signals)}
		\nl Compute $\hR=\tfrac{1}{T}\sum_{t=1}^{T}\ur[t]\ur[t]^{\dagger}$;\\
		\nl Compute $\hmLambda_{\xi}$ using \eqref{noise_approx_mean} and \eqref{approximated_covariance_noise}--\eqref{approximated_covariance_noise_last} based on $\hR$;\\
		\nl Construct $\uy$ and $\H$ according to \eqref{defofyvectorAppA0} and \eqref{widequationDefofH}, resp.;\\
		\nl Compute $\widehat{\utheta}_{\tiny{\text{ML-OWLS}}}$ via \eqref{MLbasedWLSestimate};\\
		\nl Compute $\widehat{\upsi}_{\tiny{\text{ML-OWLS}}}$ and $\widehat{\uphi}_{\tiny{\text{ML-OWLS}}}$ via \eqref{OWLS_estimates1} and \eqref{OWLS_estimates2}, resp., and denote: $\widehat{\mPsi}\triangleq\Diag\left(\widehat{\upsi}_{\tiny{\text{ML-OWLS}}}\right),\widehat{\mPhi}\triangleq\Diag\left(e^{\jmath\widehat{\uphi}_{\tiny{\text{ML-OWLS}}}}\right)$;\\
		\nl Return $\left\{\widehat{\ur}[t]\triangleq\widehat{\mPsi}^{-1}\widehat{\mPhi}^{*}\ur[t]\right\}_{t=1}^T$.
		\caption{{\bf ML-OWLS Blind Calibration Scheme} \label{Algorithm1}}}
\end{algorithm}

\section{Adaptation to an Extended Signal Model}\label{sec:extendedsignalmodel}
Before we demonstrate empirically our analytical results for model \eqref{modelequation}, we briefly present the required adaptations in order to use the proposed method for an extended signal model
\begin{equation}\label{modelequationextended}
\ur_w[t]=\ur[t]+\uw[t]=\mPsi\mPhi\ux[t]+\uw[t]\in\Cset^{M\times1},
\end{equation}
in which $\uw[t]$ denotes a possible additional additive noise vector, unaffected by the gain and phase offsets $\upsi, \uphi$. We also assume that $\uw[t]$ is independent of all the other signals involved, and is i.i.d.\ zero-mean circular CN, with a covariance matrix $\R_w\triangleq\Eset\left[\uw[t]\uw[t]^{\dagger}\right]=\sigma_w^2\I_M$. Hence, since $\uw[t]$ is independent of $\ur[t]$ and using \eqref{covariance_of_r}, the covariance matrix of $\ur_w[t]$ reads
\begin{equation}\label{covariancematrixofrw}
\mSigma\hspace{-0.01cm}\triangleq\hspace{-0.01cm}\Eset\left[\ur_w[t]\ur_w[t]^{\dagger}\right]\hspace{-0.025cm}=\hspace{-0.025cm}\mPsi\mPhi\C\mPhi^*\mPsi+\sigma_w^2\I_M\in\Cset^{M\times M}.
\end{equation}

This extended model covers a few signal models addressed in the literature, where $\uw[t]$ usually accounts for internal (e.g., thermal) receiver noise. We thus consider three cases of \eqref{modelequationextended}.\vspace{0.1cm}

\noindent\underline{\textit{Case I: Known ``Noise Floor" Level}}\\
In this case, we assume that the internal noise level, $\sigma_w^2$, is known \textit{a-priori}. This is a reasonable assumption in various cases, since the internal noise level in many receivers is (approximately) fully determined by the bandwidth of the pre-samplers filters. Thus, the Diagonally-Shifted (DS) estimate
\begin{equation}\label{meannormalizedcovestimate}
\hR_{\text{DS}}\triangleq\frac{1}{T}\sum_{t=1}^{T}{\ur_w[t]\ur_w[t]^{\dagger}}-\sigma_w^2\I_M\triangleq\hmSigma-\R_w\in\Cset^{M\times M},
\end{equation}
is an unbiased, consistent estimate of $\R$. Replacing $\hR_{\text{DS}}$ with $\hR$ everywhere in our derivation yields, for this case as well, asymptotically optimal estimates of $\upsi$ and $\uphi$ w.r.t.\ the raw data. This result follows from the same considerations presented in Subsection \ref{subsec:MLEestimation}, since $\hmSigma$ is a sufficient statistic, and $\hR_{\text{DS}}$ is an invertible function of $\hmSigma$ (since $\sigma_w^2$ is known).

\textit{Remark:} Note that in this case, the scenario is ``blind" w.r.t.\ the latent sources, their locations and $\sigma_v^2$, but since $\sigma_w^2$ is known, it may be considered (or termed) as ``semi-blind".\vspace{0.1cm}

\noindent\underline{\textit{Case II: Unknown ``Noise Floor" Level, Zero ``Interference"}}\\
In this case, we assume that the internal noise level, $\sigma_w^2$, is unknown, and $\sigma_v^2=0$ (e.g., \cite{liu2011eigenstructure}). It follows that
\begin{equation}\label{rankinequality}
\underbrace{\text{rank}\left(\R\right)=N}_{\text{num\addraAQ{b}er of sources}}<\underbrace{\text{rank}\left(\mSigma\right)=M}_{\text{num\addraAQ{b}er of sensors}}.
\end{equation}
For simplicity, we assume that $N$ is known, although, in practice, it may be estimated (e.g., via \cite{wax1985detection,kritchman2009non,weiss2019blind}). Therefore, when $M$ is known, the internal noise level $\sigma_w^2$ can be consistently estimated, e.g., via ML estimation (see \cite{wax1985detection}, Section IV, Eq.\ (13b)). Denoting this estimate as $\widehat{\sigma}_w^2$, we now define the (plug-in) ML-based DS estimate as
\begin{equation}\label{meannormalizedcovestimate2}
\hR_{\text{ML-DS}}\triangleq\hmSigma-\widehat{\sigma}_w^2\I_M\triangleq\hmSigma-\hR_w\in\Cset^{M\times M},
\end{equation}
which is a consistent estimate of $\R$. Replacing $\hR_{\text{DS}}$ with $\hR$ everywhere in our derivation yields, for this case as well, asymptotically optimal estimates of $\upsi$ and $\uphi$ w.r.t.\ the raw data. Similarly, this result follows from the same considerations presented in Subsection \ref{subsec:MLEestimation}, since $\hmSigma$ is (again) a sufficient statistic, and $\hR_{\text{ML-DS}}$ is an invertible function of $\hmSigma$.

\textit{Remark:} Note that in this case, the only deviation from a fully blind scenario is caused by the fact that the number of sources, $N$, is assumed known. However, as pointed out above, this assumption may be relaxed, as the number of sources may be consistently estimated (while still assuming $N<M-1$).\vspace{0.1cm}

\noindent\underline{\textit{Case III: Unknown ``Noise Floor"}}\\
In this case we assume that $\sigma_w^2$ is unknown, such that all the model parameters are unknown, namely a fully blind scenario. Since in this case, in general, $\text{rank}\left(\R\right)=\text{rank}\left(\mSigma\right)=M$, we propose the following non-optimal adaptation. First, note that the unbiased estimate $\hmSigma$, defined in \eqref{meannormalizedcovestimate}, is the MLE of $\mSigma$. Second, observe that \eqref{real_exact_linear} still holds for all pairs $i\neq j\in\{1,\ldots,M\}$, .i.e., for all the off-diagonal elements $\{\widehat{\Sigma}_{ij}\}$, replacing $\{\widehat{R}_{ij}\}$ in \eqref{log_equation_exact}. Observe that \eqref{imag_exact_linear}, relating to the phases, is relevant \textit{only} for the off-diagonal elements in the first place. Therefore, discarding the $M$ diagonal elements $\{\widehat{\Sigma}_{ii}\}_{i=1}^{M}$, and using (only) all the other remaining $0.5M(M-1)$ off-diagonal elements (recall that $\hmSigma$ is Hermitian), we may compactly write the \textit{reduced} linear ``correlation measurements" model
\begin{equation}\label{reduced_linear_LS_model}
\widetilde{\uy}=\widetilde{\H}\utheta+\widetilde{\uxi}\in\Rset^{M(M-1)\times 1},
\end{equation}
where $\widetilde{\uy}, \widetilde{\H}$ and $\widetilde{\uxi}$ are constructed in exactly the same way as described in Appendix \ref{AppA0}, only without including the $M$ equations associated with the diagonal elements $\{\widehat{\Sigma}_{ii}\}_{i=1}^{M}$. As long as $M(M-1)\geq K_{\theta}$, which implies $M\geq4$, the following Reduced ML-based OWLS estimate
\begin{gather}\label{reducedWLSestimate}
\widehat{\utheta}_{\tiny{\text{R-ML-OWLS}}}\triangleq\left(\widetilde{\H}^{\tps}\thmLambda_{\xi}^{-1}\widetilde{\H}\right)^{-1}\widetilde{\H}^{\tps}\thmLambda_{\xi}^{-1}\left(\widetilde{\uy}-\widehat{\widetilde{\ueta}}_{\xi}\right),
\end{gather}
is consistent, and still provides the enhancement due to its optimal weighting, where $\widehat{\widetilde{\ueta}}_{\xi}\approx\widetilde{\ueta}_{\xi}\triangleq\Eset\left[\widetilde{\uxi}\right]\in\Rset^{M(M-1)\times1}$ and $\thmLambda_{\xi}\approx\widetilde{\mLambda_{\xi}}\triangleq\Eset\left[\left(\widetilde{\uxi}-\widetilde{\ueta}_{\xi}\right)\left(\widetilde{\uxi}-\widetilde{\ueta}_{\xi}\right)^{\tps}\right]\in\Rset^{M(M-1)\times M(M-1)}$ are the ML-based estimates, computed exactly in the same manner as $\widehat{\ueta}_{\xi}$ and $\hmLambda_{\xi}$ are, resp.

Note that, although \eqref{reducedWLSestimate}---which does not use all the elements of the sufficient statistic $\hmSigma$---is not optimal (even asymptotically), the ``efficiency gap" from the performance of the exact MLE (generally) becomes negligible as the number of sensors increases since
\begin{equation}\label{numberofSensorsApproximation}
\frac{\#\text{number of discarded eq.}}{\#\text{number of total eq.}}=\frac{M}{M^2}\xrightarrow[M\rightarrow\infty]{}0.
\end{equation}
Hence, intuitively (and informally), if we assume that the ``information" regarding the unknown parameters $\upsi$ and $\uphi$ is approximately ``uniformly distributed" over all the $M^2$ potential equations (associated with the real and imaginary parts of the upper (or lower) triangular part of $\mSigma$), then the loss, in terms of the number of equations, caused by discarding the $M$ equations associated with the main diagonal of $\mSigma$, relating to gains only, becomes negligible for an array with a high number of sensors.

While this solution is no longer optimal, it still enjoys a relatively low-complexity implementation, due to its non-iterative nature (unlike, e.g., \cite{wijnholds2012statistically}), and still utilizes (most of) the implicit valuable information encapsulated in the SOS $\mSigma$, in the form of optimal weighting. Finally, note that the proposed adapted solution for this case is valid even for spatially non-white noise, namely when $\R_v$ and $\R_w$ are arbitrary (semi-positive definite) diagonal matrices.}

\section{Simulation Results}\label{sec:simulationresults}
In this section\addra{,} we consider \delra{two}\addra{three} simulation experiment\addra{s} in order to corroborate our analytical derivations by empirical results. First, we demonstrate the asymptotic optimality of the proposed estimates and the relatively substantial gain w.r.t.\ the original LS-based estimates\addra{, which commonly serve as a benchmark method}. \delra{Then}\addra{Second, we demonstrate the consistency of the modified estimate \eqref{reducedWLSestimate}, while showing at the same time that, although our method is designed for temporally i.i.d., proper sources, it in fact enables a considerable improvement even for non i.i.d.\ and/or non-proper sources, using an example of cyclostationary digital communication signals. Lastly}, we \delra{turn to }demonstrate the enhanced performance of an optimally \addra{blindly} calibrated array in DOA estimation.

\subsection{Asymptotically Optimal Blind Calibration}\label{subsec:exp1MSEs}
\begin{figure*}
	\includegraphics[width=0.98\textwidth]{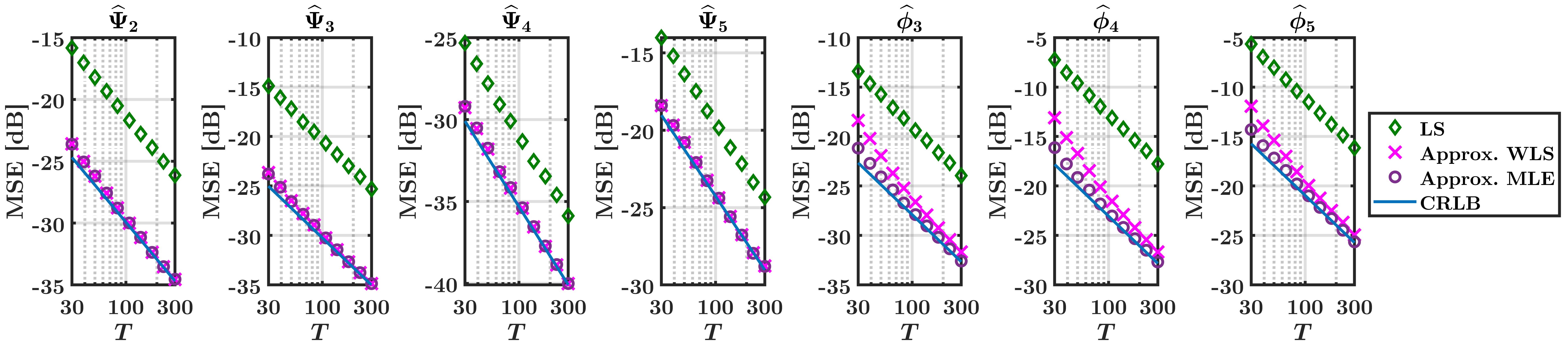}
	\caption{MSE vs.\ $T$ for SNR$=10$[dB].\delra{ Empirical results were obtained by averaging $10^4$ independent trials.} As seen, the achieved gain by the proposed estimates w.r.t.\ ordinary LS is substantial even \addra{for}\delra{in the} large sample size\addra{s}\delra{ regime}.}
	\label{fig:MSE_vs_T}\vspace{-0.3cm}
\end{figure*}
\begin{figure*}
	\includegraphics[width=0.98\textwidth]{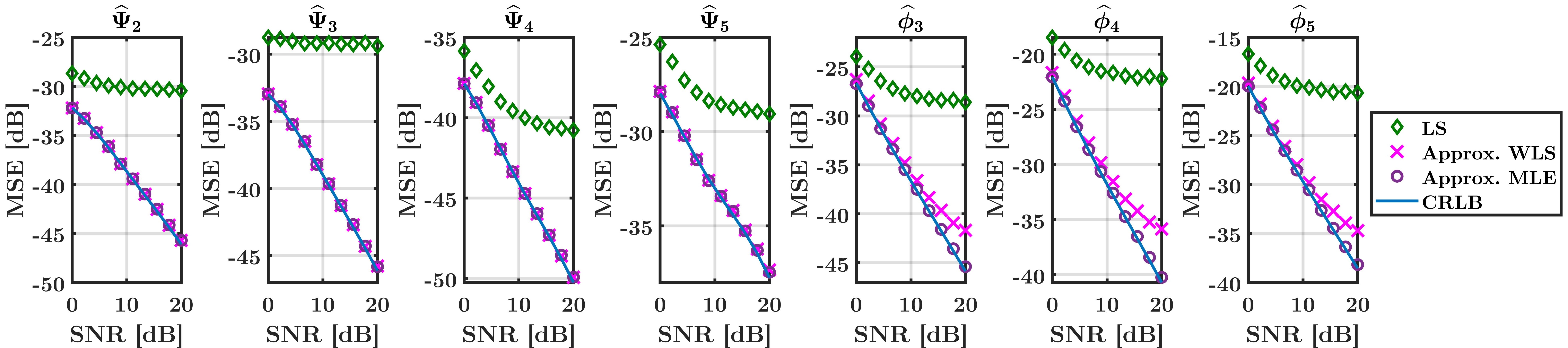}
	\caption{MSE vs.\ SNR for $T=\delra{30}\addra{75}0$.\delra{ Empirical results were obtained by averaging $10^4$ independent trials.} Evidently, the improvement w.r.t.\ the ``na\"ive" \delra{equally-weighted}\addra{ordinary} LS approach can reach more than an order of magnitude in the high SNR regime.\addra{ Our proposed estimates require fewer samples than the non-optimal (separated) WLS estimates in order to attain the CRLB in high SNRs.}}
	\label{fig:MSE_vs_SNR}\vspace{-0.3cm}
\end{figure*}
\begin{figure*}
	\includegraphics[width=0.98\textwidth]{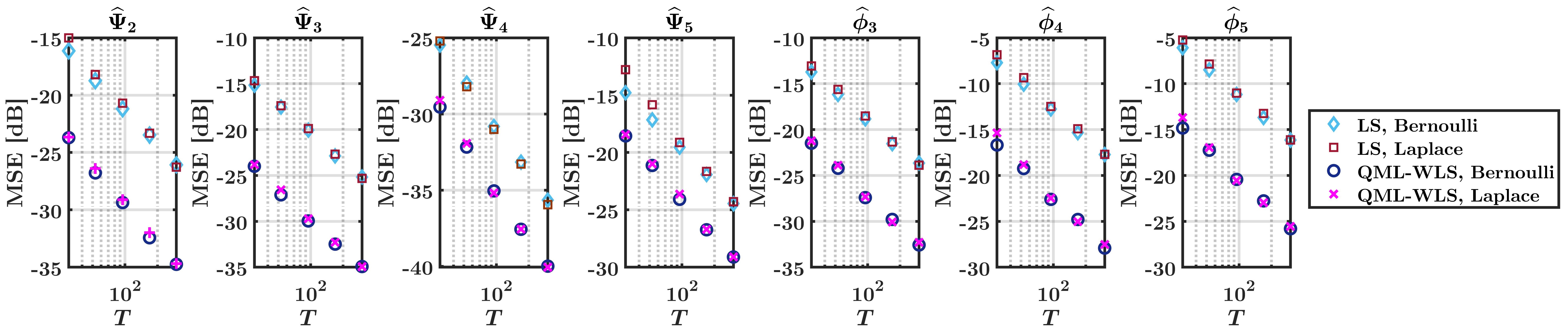}
	\caption{MSE vs.\ $T$ for SNR$=10$[dB], non-Gaussian sources and uniformly distributed noise.\delra{ Empirical results were obtained by averaging $10^4$ independent trials.} The QML-based WLS estimate is seen to be consistent and considerably better than the ``na\"ive" equally-weighted LS. Further, its robustness to signals with different fourth-order statistics is evident.}
	\label{fig:MSE_vs_T_nonGaussian}\vspace{-0.3cm}
\end{figure*}
\begin{figure*}
	\includegraphics[width=0.98\textwidth]{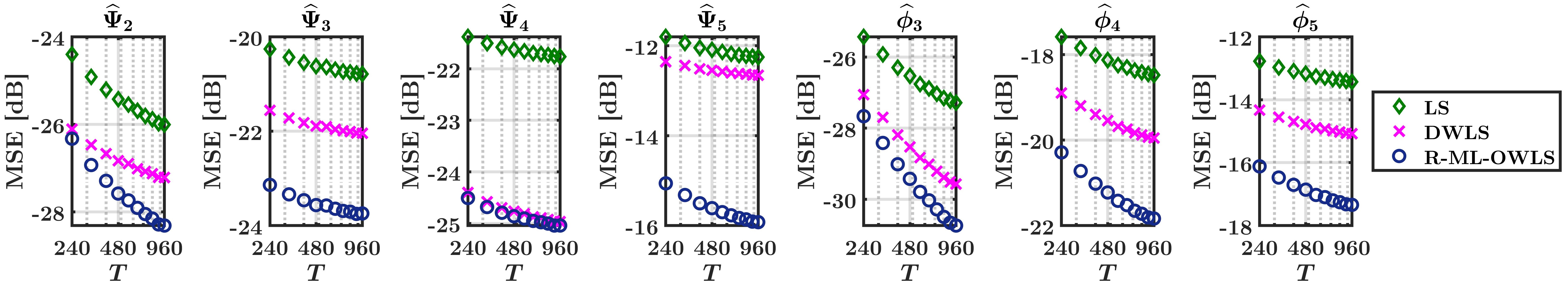}
	\caption{\addra{MSE vs.\ the sample size $T\delraAQ{=K_{f}\cdot T_{f}}$ for cyclostationary $8$-PSK OFDM and $4$-PAM communication sources with\addraAQ{ different baud rates and with} $\sigma_v^2=\sigma_w^2=0.1$, i.e., SNR$=10$[dB]. Our proposed adapted method for model \eqref{modelequationextended} still provides an improvement in the overall performance. Results are based on $5\cdot10^4$ independent trials.}}
	\label{fig:MSE_vs_T_cyclostationary}\vspace{-0.3cm}
\end{figure*}
We consider model \eqref{modelequation} in a scenario which consists of a $M=5$ elements array with half wavelength inter-element spacing (i.e., $\gamma=\lambda/2$), and $N=3$ equal power, zero-mean unit variance sources impinging from angles $\ualpha=-[35^\circ\;\delra{6}\addra{7}3^\circ\;2\delra{5}\addra{8}^\circ]^{\tps}$. The sensors\delra{'} gains and phases were set to $\upsi=[1\;1.3\;1.1\;0.7\;2.2]^{\tps}$ and $\uphi=[0^\circ\;0^\circ\;5^\circ\;11^\circ\;-8^\circ]^{\tps}$, resp., where w.l.o.g.\ we assume (throughout all Section \ref{sec:simulationresults}) that $\psi_1, \phi_1$ and $\phi_2$ are known (and serve as references). Empirical results were obtained by averaging $10^4$ independent trials.

First, we consider Gaussian signals. Fig.\ \ref{fig:MSE_vs_T} presents the MSEs obtained by $\widehat{\upsi}_{\tiny{\text{ML-OWLS}}}$ and $\widehat{\uphi}_{\tiny{\text{ML-OWLS}}}$ vs.\ $T$, where the SNR is fixed at $10$[dB]. For comparison, we also show the MSEs obtained by Paulraj and Kailath's LS estimates (ignoring the transformed measurements noise' bias and covariance),\addra{ the non-optimal WLS estimates \cite{weiss2019optimalblind}, ignoring the cross correlations \eqref{approximated_covariance_noise_last},} and the CRLB on the corresponding MSEs obtained in any unbiased joint estimation of all the unknown parameters. Similarly, Fig.\ \ref{fig:MSE_vs_SNR} presents the same quantities, however now vs.\ the SNR, where the sample size is fixed at $T=\delra{30}\addra{75}0$. As seen, the proposed estimates exhibit optimal performance, asymptotically attaining the CRLB, i.e., the asymptotic performance of the MLE based on the raw data. Notice that although this optimality is theoretically obtained only asymptotically, in practice, this asymptotic state may be reached within (only) a few dozens of samples. Moreover, the improvement w.r.t.\ ordinary LS estimation can reach more than an order of magnitude in the high SNR regime.\addra{ The improvement w.r.t.\ the non-optimal WLS estimates is mainly in phases estimation, and is reflected by the fact that for high SNRs, fewer samples are required in order to attain the CRLB.
\begin{figure}[t]
	\includegraphics[width=0.49\textwidth]{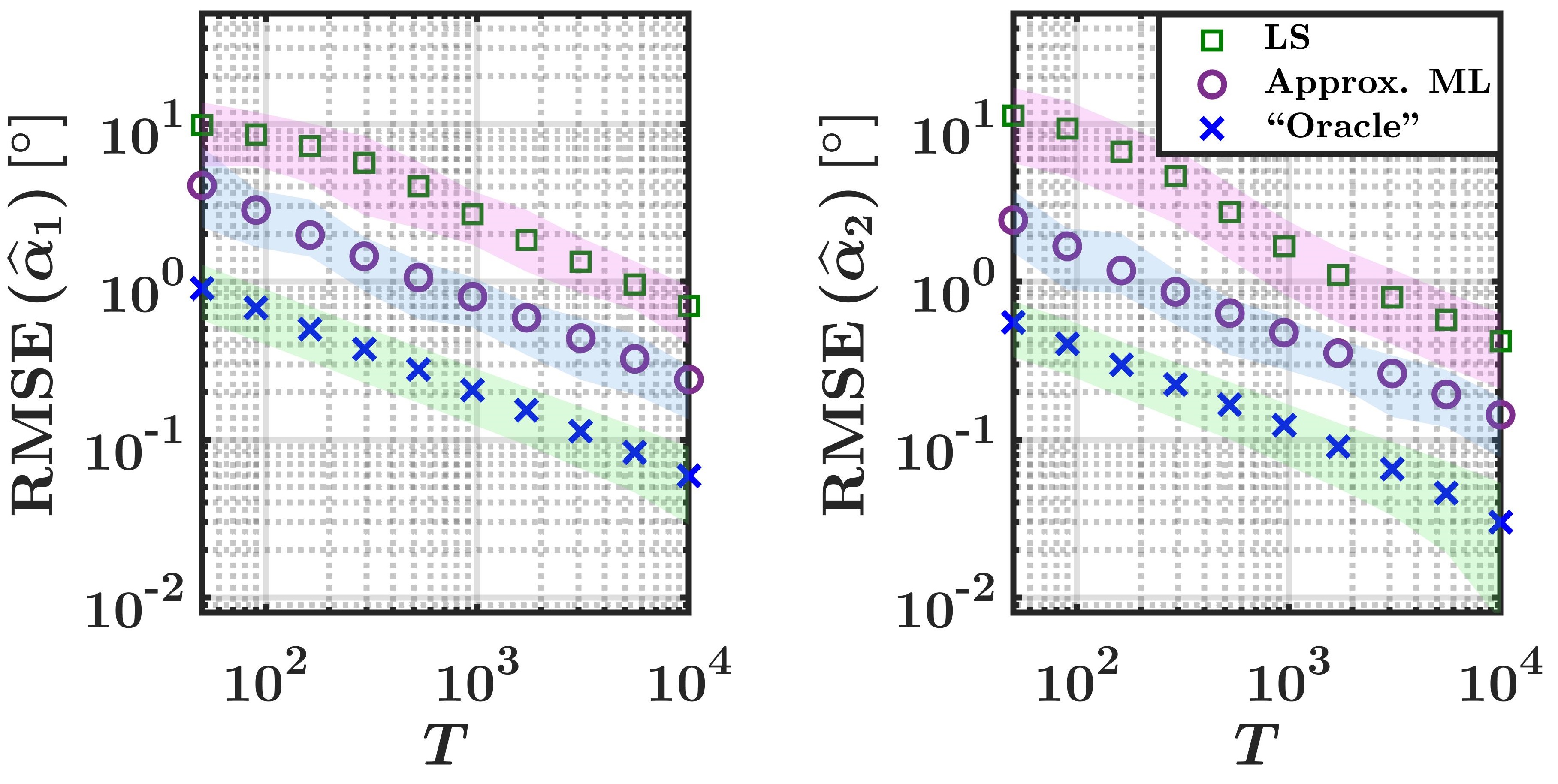}
	\caption{DOAs RMSEs $[^\circ]$ vs.\ $T$ for a fixed SNR of $0$[dB].\delra{ Empirical results were obtained by averaging $10^4$ independent trials. As evident, t}\addra{ T}he proposed \delra{(asymptotically) optimal blind calibration }scheme substantially improves the resulting\delra{ performance in DOA estimation} \addra{accuracy }w.r.t.\ ordinary LS, and is significantly closer to the \addra{accuracy}\delra{performance} attained by the perfectly calibrated array.\delra{ (a) RMSE of $\widehat{\alpha}_1$ (b) RMSE of $\widehat{\alpha}_2$}\addra{ The shaded colored areas are of the width of the respective standard deviation.}}
	\label{fig:DOA_vs_T}\vspace{-0.25cm}
\end{figure}
}
\delra{\begin{figure}[]
	\centering
	\begin{subfigure}[b]{0.24\textwidth}
		\includegraphics[width=\textwidth]{DOA1_RMSE_vs_T}
		\caption{}
		\label{fig:DOA1_vs_T}
	\end{subfigure}%
	~
	\begin{subfigure}[b]{0.24\textwidth}
		\includegraphics[width=\textwidth]{DOA2_RMSE_vs_T}
		\caption{}
		\label{fig:DOA2_vs_T}
	\end{subfigure}
	\caption{DOAs RMSEs $[^\circ]$ vs.\ $T$ for a fixed SNR of $0$[dB].\delra{ Empirical results were obtained by averaging $10^4$ independent trials.} As evident, the proposed \delra{(}asymptotically\delra{)} optimal blind calibration scheme substantially improves the resulting performance in DOA estimation w.r.t.\ ordinary LS, and is significantly closer to the performance attained by the perfectly calibrated array. (a) RMSE of $\widehat{\alpha}_1$ (b) RMSE of $\widehat{\alpha}_2$}
	\label{fig:DOA_vs_T_old}
\end{figure}
\begin{figure*}[]
	\centering
	\begin{subfigure}[b]{0.5\textwidth}
		\includegraphics[width=\textwidth]{DOA1_vs_SNR}
		\caption{}
		\label{fig:DOA1_vs_SNR}
	\end{subfigure}%
	~
	\begin{subfigure}[b]{0.5\textwidth}
		\includegraphics[width=\textwidth]{DOA2_vs_SNR}
		\caption{}
		\label{fig:DOA2_vs_SNR}
	\end{subfigure}
	\caption{DOAs RMSEs $[^\circ]$ vs.\ the SNR for a fixed sample size of  $T=\delra{3}\addra{1}000$.\delra{ Empirical results were obtained by averaging $10^4$ independent trials.} The proposed \delra{(}asymptotically\delra{)} optimal blind calibration scheme enables to reach a considerably higher accuracy level, closer to the one obtained by the ``oracle" perfectly calibrated array. (a) RMSE of $\widehat{\alpha}_1$ (b) RMSE of $\widehat{\alpha}_2$}
	\label{fig:DOA_vs_SNR_old}
\end{figure*}}

In the second part of this experiment we consider the same scenario as described above, only now with non-Gaussian signals. Specifically, the real and imaginary parts of $\uv[t]$ are mutually independent and are (equally) uniformly distributed with zero-mean. For the sources, we consider the Bernoulli (with a parameter $p=0.5$) and Laplace distributions (again, for the mutually independent real and imaginary parts), which were chosen as representatives of both platykurtic and leptokurtic distributions, characterized by less and more frequent occurrences of extreme outliers, resp., than the Gaussian distribution. In both cases, the sources were normalized to have zero mean and unit variance. Fig.\ \ref{fig:MSE_vs_T_nonGaussian} is the same as Fig.\ \ref{fig:MSE_vs_T}, only now it demonstrates the performance of the QML-based WLS estimate, which, in practice, is computed exactly as the one presented in the first part of this experiment, but is no longer considered the ML-based OWLS, as explained in Subsection \ref{sec:QMLcalibration}. Evidently, this estimate is also consistent and performs better than the ordinary LS estimate. Furthermore, it is seen that the fourth-order statistics of the received signals effectively have little influence asymptotically. For different SNR values, a similar trend \delra{to the one shown}\addra{as} in Fig.\ \ref{fig:MSE_vs_SNR} is obtained\footnote{This, of course, was validated by simulations.}.
\addra{\subsection{Cyclostationary Digital Communication Sources }\label{subsec:cyclostationary}
We now consider the extended model \eqref{modelequationextended} in a similar, yet different scenario relative to Subsection \ref{subsec:exp1MSEs}. The setting is identical w.r.t.\ the receiver (/system) parameters $M,\gamma,\upsi$ and $\uphi$. However, in this experiment the $N=3$ sources are cyclostationary digital communication signals \cite{gardner1994introduction}, emitted from angles $\ualpha=-[45^\circ\;52^\circ\;13^\circ]^{\tps}$. Notice that here, the first two sources are less radially separated. Each source is constructed by concatenating $K_f$ frames, each of length $T_f$ samples. More specifically, the $n$-th source is given by
\begin{equation}\label{cyclostationarysource}
s_n[t]=\sum_{k=0}^{K_f-1}{f^{(k)}_n[t-k\cdot T_f]}\in\Cset,
\end{equation}
where each individual frame is defined by
\begin{equation}\label{framedefinition}
f^{(k)}_n[t]=\begin{cases}
\delta_{t1}-\delta_{t2},& 0\leq t< T_{\text{sync}}\\
\vartheta^{(k)}_n[t-T_{\text{sync}}],& T_{\text{sync}}\leq t<T_f
\end{cases},
\end{equation}
such that the first $T_{\text{sync}}$ samples are synchronization guard intervals, and the following $T_f-T_{\text{sync}}$ samples are dedicated to the information symbols. Here, we set $T_f=40, T_{\text{sync}}=8$, such that each symbols packet $\vartheta^{(k)}_n[t]$ is $2^5$ samples long.\addraAQ{
	
In order to simulate signals with different baud rates and frame-synchronization, we applied different ``time-stretch" factors to $s_2[t]$ and $s_3[t]$, replacing these signals with $s_2\left[\left\lfloor(t-1)/2\right\rfloor\right]$ and $s_3\left[\left\lfloor(t-2)/3\right\rfloor\right]$, resp. (where $\left\lfloor \tau \right\rfloor$ denotes the ``floor" operator, namely, the largest integer smaller than or equal to $\tau$). Consequently, the complete observation time $T=K_f\cdot T_f$ contains $K_f$ frames of $s_1[t]$, but only (approximately) $K_f/2$ and $K_f/3$ frames of $s_2[t]$ and $s_3[t]$, resp.}

Fig.\ \ref{fig:MSE_vs_T_cyclostationary} presents the MSEs obtained by $\widehat{\upsi}_{\tiny{\text{R-ML-OWLS}}}$ and $\widehat{\uphi}_{\tiny{\text{R-ML-OWLS}}}$ (extracted from \eqref{reducedWLSestimate}) vs.\ $T\delraAQ{(=K_{f}\cdot T_{f})}$, where the first two sources' symbols packets are unit variance 8 Phase Shift Keying ($8$-PSK) Orthogonal Frequency Division Multiplexed (OFDM) signals\delraAQ{\addraAQ{\footnote{\addraAQ{Note that the lengths of the OFDM packets $\vartheta^{(k)}_1[t]$ and $\vartheta^{(k)}_2[t]$ are $32=2^5$ and $64=2^6$, and are therefore FFT compatible, as required.}}}}, the third source's symbols packets are (real-valued) zero-mean, unit variance $4$-level Pulse Amplitude Modulated ($4$-PAM) signals, and $\sigma_v^2=\sigma_w^2=0.1$ fixed (i.e., $\text{SNR}=10$[dB]). All symbols were equiprobable, and were drawn independently. For comparison, we also show the MSEs obtained by Paulraj and Kailath's ordinary LS estimates and Liu \etal's Diagonal WLS (DWLS) \cite{liu2010precision}. Clearly, our proposed adapted method still offers a considerable performance improvement, even for non i.i.d., non stationary sources.
}
\begin{figure}[t]
	\includegraphics[width=0.485\textwidth]{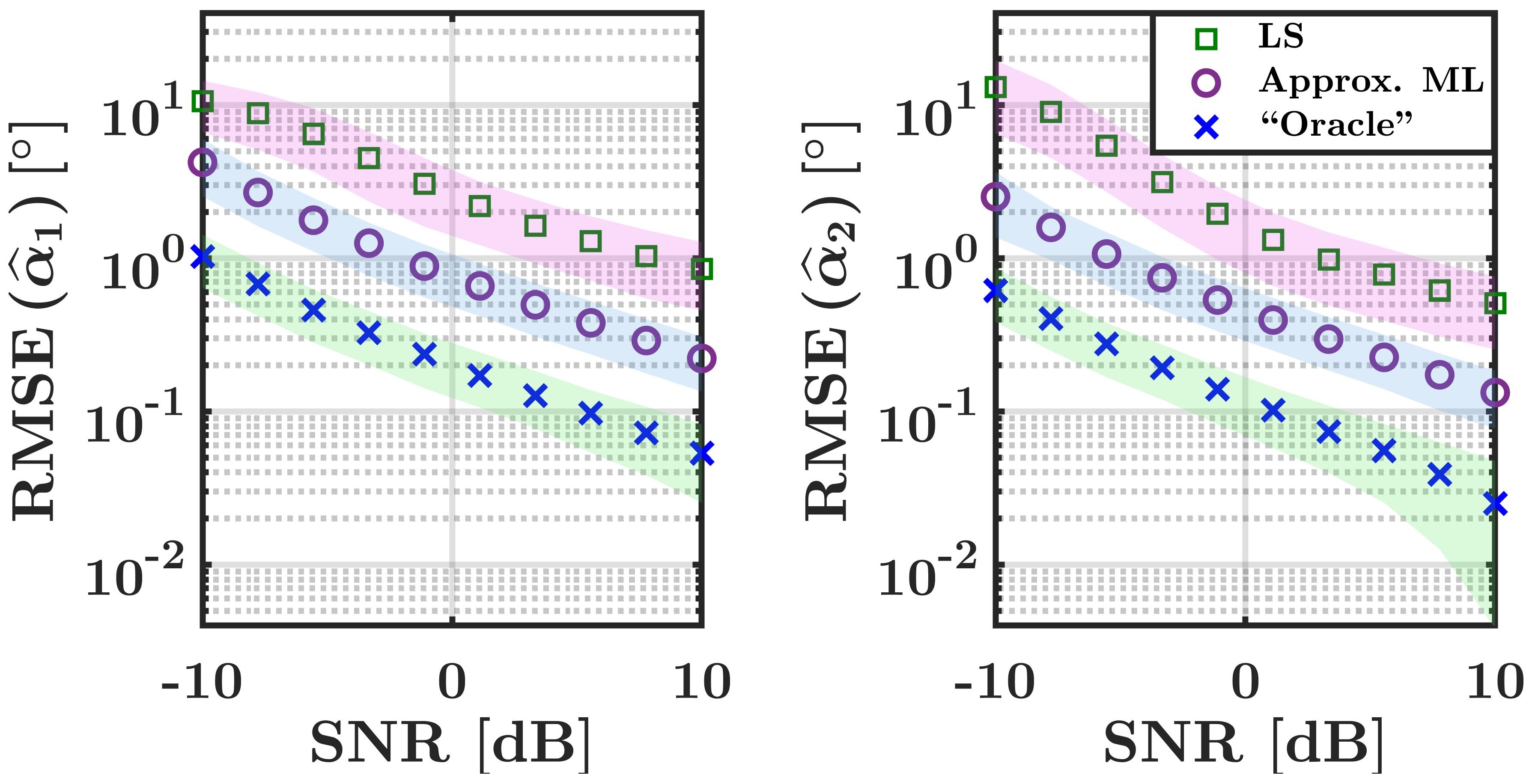}
	\caption{DOAs RMSEs $[^\circ]$ vs.\ the SNR for a fixed sample size of  $T=\delra{3}\addra{1}000$.\delra{ Empirical results were obtained by averaging $10^4$ independent trials.} The proposed \delra{(asymptotically) optimal} blind calibration scheme enables\delra{ to reach} a considerably higher accuracy level, closer to the one \delra{obtained}\addra{attained} by the ``oracle" perfectly calibrated array.\delra{ (a) RMSE of $\widehat{\alpha}_1$ (b) RMSE of $\widehat{\alpha}_2$}\addra{ The shaded colored areas are of the width of the respective standard deviation.}}
	\label{fig:DOA_vs_SNR}\vspace{-0.5cm}
\end{figure}
\subsection{DOA Estimation via MUSIC}\label{subsec:exp2DOA}
We consider a similar scenario with an identical array (and the same offsets $\upsi,\uphi$), only now we assume that two ($N=2$) zero-mean unit variance Gaussian sources are emitted from (unknown) angles $\ualpha=-[35^\circ\;\delra{6}\addra{7}3^\circ]^{\tps}$, where $N$ is assumed to be known, and the goal here is DOA estimation, which is done via the MUSIC algorithm. For a fixed SNR level of $0$[dB], Fig\delra{s}.\ \delra{\ref{fig:DOA1_vs_T} and \ref{fig:DOA2_vs_T}}\addra{\ref{fig:DOA_vs_T}} present\addra{s} the Root MSE (RMSE) of the DOAs estimates, $\widehat{\alpha}_1, \widehat{\alpha}_2$, vs.\ the sample size of three different post-calibration MUSIC estimates, corresponding to LS blind calibration, \delra{(}asymptotically\delra{)} optimally blind calibration and ``oracle" perfect calibration. Here as well, already for a relatively small sample size (in the order of the number of unknown parameters), a significant improvement in the resulting accuracy is demonstrated, reaching up to \addra{nearly }an order of magnitude\addra{ for low sample sizes} w.r.t.\ the \delra{ordinary }LS calibration-based estimates' RMSEs. A similar trend of enhanced accuracy is shown in Fig\delra{s}.\ \delra{\ref{fig:DOA1_vs_SNR} and \ref{fig:DOA2_vs_SNR}}\addra{\ref{fig:DOA_vs_SNR}}, presenting the DOAs estimates' RMSEs vs.\ the SNR for a fixed sample size of $T=\delra{3}\addra{1}000$.

\section{Conclusion}\label{sec:conclusion}
In the context of ULAs, we presented an asymptotically optimal blind calibration scheme for narrowband Gaussian signals. Based on the Toeplitz structure of the observations' covariance matrix and on asymptotic approximations, we derived OWLS estimates, which were shown to be \delra{(}asymptotically\delra{)} equivalent to the MLEs of the sensors\delra{'} gain and phase offsets in joint ML estimation of all the unknown parameters. Additionally, we derived the CRLB on the MSEs of any unbiased estimate thereof, which are attained asymptotically by our proposed estimates. Our analytical results and the significant performance gain were demonstrated in simulation experiments, where we also showed the resulting enhanced accuracy in a post-calibration DOAs estimation task.

For non-Gaussian signals, the proposed estimates serve as QML estimates, which are still asymptotically optimal w.r.t.\ the empirical covariance matrix, but, in general, are no longer the MLEs w.r.t.\ the raw data. Nevertheless, in comparison to Paulraj and Kailath's ordinary LS estimates, these estimates still exhibit a considerable improvement in the resulting performance, as demonstrated empirically in simulations, eventually enabling higher accuracy in other post-calibration\delra{ estimation} procedures\delra{.}\addra{, such as DOA estimation}

\delra{We note that the principles presented in this work may also be applied in order to derive (a similar, yet different) blind calibration scheme of ULAs in measurement models for which the additive noise term is not affected by the sensor\delra{'}s offsets (giving rise to somewhat different expressions for the optimal estimates).}

\begin{figure*}[t]
\addra{
	\begin{equation*}\label{widequationDefofH}
	\H\triangleq\begin{bmatrix}
	\text{lvec}\left(\H^{\psi}[1]\right) \;\; \cdots \;\; \text{lvec}\left(\H^{\psi}[M]\right) \;\; \O \;\; \text{lvec}\left(\H^{\rho}[1]\right) \;\; \cdots \;\; \text{lvec}\left(\H^{\rho}[M]\right) \;\; \O\\
	\O \;\; \text{uvec}\left(\H^{\phi}[1]\right) \;\; \cdots \;\; \text{uvec}\left(\H^{\phi}[M]\right) \;\; \O \;\; \text{uvec}\left(\H^{\iota}[1]\right) \;\; \cdots \;\; \text{uvec}\left(\H^{\iota}[M]\right)\\
	\end{bmatrix} \in \Rset^{M^2 \times M^2}\tag{54}
	\end{equation*}
	\hrulefill}
\end{figure*}

\appendices
\addra{\section{Construction of $\uy, \H$ and $\uxi$}\label{AppA0}
Note first (from \eqref{real_exact_linear}, \eqref{imag_exact_linear}) that $\log\left(\hR\right)$ can be expressed as the following linear combination of the elements of $\utheta$:
\begin{equation}\label{AppA0eq1}
\begin{gathered}
\log\left(\hR\right)\triangleq\uhmu+\jmath\uhnu=\sum_{m=1}^{M}{\H^{\psi}[m]}\tpsi_m+\jmath\sum_{m=1}^{M}{\H^{\phi}[m]}\phi_m\\
+\sum_{m=1}^{M}{\H^{\rho}[m]}\rho_m+\jmath\sum_{m=1}^{M}{\H^{\iota}[m]}\iota_m+\dbaruvarep+\jmath\cdot\dbaruepsilon
\end{gathered}
\end{equation}
with the following real-valued $M\times M$ matrices: $\uhmu, \dbaruvarep, \uhnu$ and $\dbaruepsilon$ consist of the elements $\widehat{\mu}_{ij}, \varepsilon_{ij}, \widehat{\nu}_{ij}$ and $\epsilon_{ij}$ (resp.) as defined in \eqref{real_exact_linear}, \eqref{imag_exact_linear}, resp.; and for all $i,j,m\in\{1,\ldots,M\}$,
\begin{equation}\label{AppA0eq2}
\begin{gathered}
H^{\psi}_{ij}[m]\triangleq\delta_{im}+\delta_{jm},\quad H^{\phi}_{ij}[m]\triangleq\delta_{im}+\delta_{jm},\\
H^{\rho}_{ij}[m]=H^{\iota}_{ij}[m]=\delta_{(i-j+1)m}.
\end{gathered}
\end{equation}
Now define the operators $\text{lvec}(\cdot)$ and $\text{uvec}(\cdot)$, which vectorise the lower-triangular part (including the diagonal) and strictly-upper-triangular part (excluding the diagonal) of their square matrix argument (resp.). Namely, for any $M\times M$ matrix $\A$,
\begin{align}
\text{lvec}(\A)\triangleq&\left[A_{11}\;A_{21}\;\cdots\;A_{M1}\;A_{22}\;A_{32}\;\cdots\right.\\
&\left.A_{M2}\;\cdots\;A_{M(M-1)}\;A_{MM}\right]\in\Cset^{0.5M(M+1)\times1},\nonumber\\
\text{uvec}(\A)\triangleq&\left[A_{12}\;A_{13}\;A_{23}\;A_{14}\;\cdots\;A_{34}\;\cdots\right.\\
&\left.A_{1M}\;A_{2M}\;\cdots\;A_{(M-1)M}\right]\in\Cset^{0.5M(M-1)\times1}.\nonumber
\end{align}
Using these operators we now construct:
\begin{align}\label{defofyvectorAppA0}
\uy&\triangleq\left[\text{lvec}^{\tps}\left(\Re\left\{\log\left(\hR\right)\right\}\right)\;\;\text{uvec}^{\tps}\left(\Im\left\{\log\left(\hR\right)\right\}\right)\right]^{\tps}\nonumber\\
&=\left[\text{lvec}^{\tps}\left(\uhmu\right)\;\;\text{uvec}^{\tps}\left(\uhnu\right)\right]^{\tps}\in\Rset^{M^2\times 1},
\end{align}
so that with $\H$ defined in \eqref{widequationDefofH} at the top of the page (where in the upper block $\O$ denotes a $0.5M(M+1)\times M$ all-zeros matrix and in the lower block $\O$ denotes a $0.5M(M-1)\times M$ all-zeros matrix), and
\begin{equation}\label{measurementerrorvecAppA0}
\setcounter{equation}{55}
\uxi\triangleq\begin{bmatrix}
\text{lvec}\left(\dbaruvarep\right) \\ \text{uvec}\left(\dbaruepsilon\right)
\end{bmatrix}=\begin{bmatrix}
\uvarep \\ \uepsilon
\end{bmatrix}\in\Rset^{M^2 \times 1}
\end{equation}
(where $\uvarep\in\Rset^{0.5M(M+1)\times 1}$ and $\uepsilon\in\Rset^{0.5M(M-1)\times 1}$ were defined below \eqref{linear_LS_model}), we obtain the desired relation $\uy=\H\utheta+\uxi$.

}

\section{Computation of the Noise Covariance Matrix}\label{AppA}
Our goal here is to obtain \delra{(}approximated\delra{)} closed-form expressions for the elements of the covariance matrix $\mLambda_{\xi}$. We begin with the computation of
\begin{equation}\label{noise_xi_covariance_block}
\Eset\left[\uxi\uxi^{\tps}\right]= \begin{bmatrix}
\vspace{-0.3cm}\\
\Eset\left[\uvarep\uvarep^{\tps}\right] & \Eset\left[\uvarep\uepsilon^{\tps}\right]\\
\vspace{-0.25cm}\\
\Eset\left[\uepsilon\uvarep^{\tps}\right] & \Eset\left[\uepsilon\uepsilon^{\tps}\right]\\
\vspace{-0.3cm} \end{bmatrix} \in \Rset^{M^2 \times M^2}.
\end{equation}
As seen from \eqref{log_equation_exact2} and \eqref{approximated_noise}, we have
\begin{equation}\label{complexnoise}
\zeta_{ij}=\varepsilon_{ij}+\jmath\cdot\epsilon_{ij}\approx\frac{\ep_{ij}}{R_{ij}} - \frac{\ep_{ij}^2}{2R_{ij}^2}, \; \; \forall i,j\in\{1,\ldots,M\}.
\end{equation}
Starting with the elements of the upper-left block $\Eset\left[\uvarep\uvarep^{\tps}\right]$, it may be easily shown that for any $z_1,z_2\in\Cset$
\begin{equation}\label{identity1}
\Eset\left[\Re\{z_1\}\Re\{z_2\}\right]=0.5\cdot\Re\left\{\Eset\left[z_1z_2^*\right]+\Eset\left[z_1z_2\right]\right\}.
\end{equation}
Hence, neglecting fourth-order noise terms yields
\begin{equation}\label{identity1used}
\Eset\left[\varepsilon_{ij}\varepsilon_{k\ell}\right]\approx0.5\cdot\Re\left\{\frac{\Eset\left[\ep_{ij}\ep_{k\ell}^*\right]}{R_{ij}R_{k\ell}^*}+\frac{\Eset\left[\ep_{ij}\ep_{k\ell}\right]}{R_{ij}R_{k\ell}}\right\},
\end{equation}
so we may concentrate on $\Eset\left[\ep_{ij}\ep_{k\ell}^*\right]$ and $\Eset\left[\ep_{ij}\ep_{k\ell}\right]$, the covariances and pseudo-covariances of $\{\ep_{ij}\}$. Thus,
\begin{gather}
\Eset\left[\ep_{ij}\ep_{k\ell}^*\right]=\Eset\left[\widehat{R}_{ij}\widehat{R}^*_{k\ell}\right]-R_{ij}R^*_{k\ell}\nonumber\\
=\frac{1}{T^2}\sum_{t_1,t_2=1}^{T}{\Eset\left[r_i[t_1]r_j^*[t_1]r_k^*[t_2]r_\ell[t_2]\right]}-R_{ij}R^*_{k\ell}.\label{firststep}
\end{gather}
Using the circularity of $\ur[t]$, which implies
\begin{gather}\label{circularity_of_r}
\Eset\left[r_i[t_1]r_j[t_2]\right]=\Eset\left[r_i^*[t_1]r^*_j[t_2]\right]=0,\nonumber\\
\forall i,j\in\{1,\ldots,M\}, \forall t_1,t_2\in\{1,\ldots,T\},
\end{gather}
we may write the summand in \eqref{firststep} as
\begin{multline}\label{fourth_moment_formula}
\Eset\left[r_i[t_1]r_j^*[t_1]r_k^*[t_2]r_\ell[t_2]\right]=\\
\begin{cases}
\kappa_r[i,j,\ell,k]+R_{ij}R_{k\ell}^*+R_{ik}R_{j\ell}^*, & t_1=t_2\\
R_{ij}R_{k\ell}^*, & t_1\ne t_2
\end{cases},
\end{multline}
where $\kappa_r[i,j,k,\ell]\triangleq\text{cum}(r_i[t],r_j^*[t],r_k[t],r^*_\ell[t])$ denotes the fourth-order joint cumulant of its arguments. Using the fact \delra{the}\addra{that} $\{\ur[t]\}_{t=1}^T$ are all i.i.d.\ circular CN, by applying Isserlis' theorem \cite{isserlis1918formula}, it follows that $\kappa_r[i,j,k,\ell]=0$ for all $i,j,k,\ell\in\{1,\ldots,M\}$. Accordingly, substituting \eqref{fourth_moment_formula} into \eqref{firststep}, and repeating for $\Eset\left[\ep_{ij}\ep_{k\ell}\right]$ with exactly the same technique, we obtain after simplification
\begin{gather}\label{errorcovariance}
\Eset\left[\ep_{ij}\ep_{k\ell}^*\right]=\frac{1}{T}R_{ik}R^*_{j\ell}, \quad \Eset\left[\ep_{ij}\ep_{k\ell}\right]=\frac{1}{T}R_{i\ell}R^*_{jk},
\end{gather}
for all $i,j,k,\ell\in\{1,\ldots,M\}$. \addra{Note that \eqref{errorcovariance} implies that the estimations errors $\{\ep_{ij}\}$ are non-circular, in contrast to the measured signals $\{r_i[t]\}$. }Now, substituting \eqref{errorcovariance} into \eqref{identity1used}, we obtain (for all $i,j,k,\ell\in\{1,\ldots,M\}$)
\begin{equation}\label{approximated_covariance_noise_appendix}
\Eset\left[\varepsilon_{ij}\cdot\varepsilon_{k\ell}\right]\approx\frac{1}{T}\cdot0.5\cdot\Re\left\{\frac{R_{ik}R^*_{j\ell}}{R_{ij}R^*_{k\ell}}+\frac{R_{i\ell}R^*_{jk}}{R_{ij}R_{k\ell}}\right\}.
\end{equation}

Similarly, notice that for any $z_1,z_2\in\Cset$ we also have
\begin{align}
\Eset\left[\Im\{z_1\}\Im\{z_2\}\right]&=0.5\cdot\Re\left\{\Eset\left[z_1z_2^*\right]-\Eset\left[z_1z_2\right]\right\},\label{identity2}\\
\Eset\left[\Re\{z_1\}\Im\{z_2\}\right]&=0.5\cdot\Im\left\{\Eset\left[z_1z_2\right]-\Eset\left[z_1z_2^*\right]\right\},\label{identity3}
\end{align} 
and using \eqref{errorcovariance}, which we have already obtained, we immediately have (for all $i,j,k,\ell\in\{1,\ldots,M\}$)
\begin{align}\label{final_approximated_covariance_terms_appendix}
&\Eset\left[\epsilon_{ij}\cdot\epsilon_{k\ell}\right]\approx\frac{1}{T}\cdot0.5\cdot\Re\left\{\frac{R_{ik}R^*_{j\ell}}{R_{ij}R^*_{k\ell}}-\frac{R_{i\ell}R^*_{jk}}{R_{ij}R_{k\ell}}\right\},\\
&\Eset\left[\varepsilon_{ij}\cdot\epsilon_{k\ell}\right]\approx\frac{1}{T}\cdot0.5\cdot\Im\left\{\frac{R_{i\ell}R^*_{jk}}{R_{ij}R_{k\ell}}-\frac{R_{ik}R^*_{j\ell}}{R_{ij}R^*_{k\ell}}\right\},\label{second_final_term}
\end{align}
which are the elements of all the other block matrices $\Eset\left[\uepsilon\uepsilon^{\tps}\right], \Eset\left[\uvarep\uepsilon^{\tps}\right]=\left(\Eset\left[\uepsilon\uvarep^{\tps}\right]\right)^{\tps}$ assembling \eqref{noise_xi_covariance_block}.

Using the pseudo-covariance of $\{\ep_{ij}\}$ given in \eqref{errorcovariance}, we have
\begin{multline}\label{noiseexpectation_appendix}
\Eset\left[\zeta_{ij}\right]\approx\frac{\Eset\left[\ep_{ij}\right]}{R_{ij}} - \frac{\Eset\left[\ep_{ij}^2\right]}{2R_{ij}^2}=-\frac{1}{2T}\;\delra{\Rightarrow}\addra{\quad\Longrightarrow}\\
\Eset\left[\varepsilon_{ij}\right]\approx-\frac{1}{2T},\;\Eset\left[\epsilon_{ij}\right]\approx0,\;\forall i,j\in\{1,\ldots,M\},
\end{multline}
so that 
\begin{equation}\label{noise_approx_mean_appendix}
\ueta_{\xi}\approx-\frac{1}{2T}\cdot\left[\uon^{\tps}_{0.5M(M+1)}\; \uo^{\tps}_{0.5M(M-1)}\right]^{\tps}.
\end{equation}
Now, recall that $\mLambda=\Eset\left[\uxi\uxi^{\tps}\right]-\ueta_{\xi}\ueta_{\xi}^{\tps}$, thus we have obtained in \eqref{approximated_covariance_noise_appendix}, \eqref{final_approximated_covariance_terms_appendix}--\eqref{second_final_term} and \eqref{noise_approx_mean_appendix} approximated closed-form expressions for all the elements of the covariance matrix $\mLambda$. 

\bibliography{Bibfile}
\bibliographystyle{unsrt}

\end{document}